\documentclass[fleqn,usenatbib]{mnras}

\usepackage{newtxtext,newtxmath}
\usepackage[T1]{fontenc}

\DeclareRobustCommand{\VAN}[3]{#2}
\let\VANthebibliography\thebibliography
\def\thebibliography{\DeclareRobustCommand{\VAN}[3]{##3}\VANthebibliography}

\usepackage{graphicx}	% Including figure files
\usepackage{adjustbox}  % For the orcid ids
\usepackage{amsmath}	% Advanced maths commands
\usepackage{wasysym}    % Extra astronomy symbols
\usepackage{bm}         % Bold maths
\usepackage{xspace}
\usepackage{soul}
\makeatletter % Workaround to only color the year for cite commands
  \patchcmd{\NAT@citex}
    {\@citea\NAT@hyper@{%
      \NAT@nmfmt{\NAT@nm}%
      \hyper@natlinkbreak{\NAT@aysep\NAT@spacechar}{\@citeb\@extra@b@citeb}%
      \NAT@date}}
    {\@citea\NAT@nmfmt{\NAT@nm}%
    \NAT@aysep\NAT@spacechar\NAT@hyper@{\NAT@date}}{}{}

  \patchcmd{\NAT@citex}
    {\@citea\NAT@hyper@{%
      \NAT@nmfmt{\NAT@nm}%
      \hyper@natlinkbreak{\NAT@spacechar\NAT@@open\if*#1*\else#1\NAT@spacechar\fi}%
        {\@citeb\@extra@b@citeb}%
      \NAT@date}}
    {\@citea\NAT@nmfmt{\NAT@nm}%
    \NAT@spacechar\NAT@@open\if*#1*\else#1\NAT@spacechar\fi\NAT@hyper@{\NAT@date}}
    {}{}
\makeatother

\newcommand\Msun{\text{M}_{\astrosun}} % requires the wasysym package
\newcommand\HI{\ion{H}{I}\xspace} % neutral hydrogen
\newcommand\HII{\ion{H}{II}\xspace} % ionized hydrogen
\newcommand\arepo{\mbox{\textsc{arepo}}\xspace}
\newcommand\areport{\mbox{\textsc{arepo-rt}}\xspace}
\newcommand\thesan{\mbox{\textsc{thesan}}\xspace}
\newcommand\thesanone{\mbox{\textsc{thesan-1}}\xspace}
\newcommand\thesandarkone{\mbox{\textsc{thesan-dark-1}}\xspace}
\newcommand\thesantwo{\mbox{\textsc{thesan-2}}\xspace}
\newcommand\thesandarktwo{\mbox{\textsc{thesan-dark-2}}\xspace}

\newcommand\orcid[1]{\href{http://orcid.org/#1}{\adjustbox{trim={-.15\width} {0\height} {-.15\width} {0\height},clip}{\includegraphics[height=10pt]{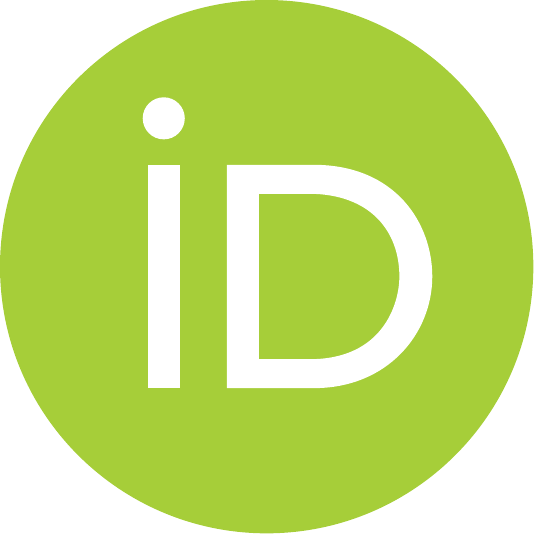}}}}
\title[Reionization in the Local Group]{The \textsc{thesan} project: environmental drivers of Local Group reionization}

\author[Y.~Zhao et al.]{%
Yu~Zhao\orcid{0000-0001-6768-9947},$^{1,2}$\thanks{E-mail: \href{mailto:yzhao112@usc.edu}{yzhao112@usc.edu}}
Aaron~Smith\orcid{0000-0002-2838-9033},$^{3}$
Rahul~Kannan\orcid{0000-0001-6092-2187},$^{4}$
Enrico~Garaldi\orcid{0000-0002-6021-7020},$^{5}$
Hui~Li\orcid{0000-0002-1253-2763},$^{6}$
\newauthor
Mark~Vogelsberger\orcid{0000-0001-8593-7692},$^{7,8}$
Andrew~Benson\orcid{0000-0001-5501-6008}$^{2}$
and
Lars~Hernquist$^{9}$
\\%
$^{1}$Department of Physics and Astronomy, University of Southern California, Los Angeles,
CA 90007, USA \\%
$^{2}$Carnegie Observatories, 813 Santa Barbara Street, Pasadena, CA 91101, USA\\%
$^{3}$Department of Physics, The University of Texas at Dallas, Richardson, Texas 75080, USA \\%
$^{4}$Department of Physics and Astronomy, York University, 4700 Keele Street, Toronto, ON M3J 1P3, Canada \\%
$^{5}$Kavli IPMU (WPI), UTIAS, The University of Tokyo, Kashiwa, Chiba 277-8583, Japan \\%
$^{6}$Department of Astronomy, Tsinghua University, Beijing 100084, People’s Republic of China \\%
$^{7}$Department of Physics, Kavli Institute for Astrophysics and Space Research, Massachusetts Institute of Technology, Cambridge, MA 02139, USA \\%
$^{8}$The NSF AI Institute for Artificial Intelligence and Fundamental Interactions, Massachusetts Institute of Technology, Cambridge MA 02139, USA \\%
$^{9}$Center for Astrophysics $\vert$ Harvard $\&$ Smithsonian, 60 Garden Street, Cambridge, MA 02138, USA%
}

\date{Accepted XXX. Received YYY; in original form ZZZ}

\pubyear{2025}

\begin{document}
\label{firstpage}
\pagerange{\pageref{firstpage}--\pageref{lastpage}}
\maketitle

\begin{abstract}
  The timing of cosmic reionization across Local Group (LG) analogues provides insights into their early histories and surrounding large-scale structure. Using the radiation-hydrodynamic simulation \thesanone and its dark matter-only counterpart \thesandarkone, we track the reionization histories of all haloes, including 224 LG analogues within the proximity of any of the $20$ Virgo-like clusters with halo masses above $10^{14}\,\Msun$ at $z = 0$ and their environments. The statistically controlled samples quantify how the reionization redshift ($z_\text{reion}$) correlates with halo mass, local overdensity, and present-day pair properties.
  Even at fixed mass, haloes in denser regions ionize earlier, and increasing the overdensity smoothing scale systematically suppresses small-scale structure, including local variations and environmental gradients in $z_\text{reion}$. Virgo-like clusters accelerate reionization in their surroundings out to $\sim$5--10\,cMpc, beyond which local overdensity again becomes the dominant factor. Within LG pairs, reionization timing offsets reach up to $\sim$150\,Myr and correlate with present-day halo separation, reflecting sensitivity to large-scale structure rather than mass ratio in driving asynchronous reionization. The results support an extreme inside-out picture where clustered sources rapidly ionize their immediate neighborhoods, while lower-density regions self-ionize later and voids wait for external homogenization. These links between environment and reionization timing explain the influence of protoclusters and help interpret fossil records in LG dwarfs around the Milky Way. For Milky Way analogues, we find a reionization redshift as early (late) as $z_{\rm reion} = 12.7^{+2.0}_{-1.7}$ ($8.88^{+0.66}_{-0.70}$) when considered on 125\,ckpc (500\,ckpc) scales, with LG analogues following an inside-out reionization picture.
\end{abstract}


%%%%%%%%%%%%%%%%%%%%%%%%%%%%%%%%%%%%%%%%%%%%%%%%%%

%%%%%%%%%%%%%%%%% BODY OF PAPER %%%%%%%%%%%%%%%%%%

\section{Introduction}
The Epoch of Reionization (EoR) marks a pivotal phase in the evolution of the Universe, representing its final major phase transition. Following the epoch of recombination at $z \approx 1100$, cosmological expansion cooled the Universe sufficiently for free electrons and protons to combine into neutral atomic hydrogen. Within the first billion years after the Big Bang, these conditions were gradually reversed as the formation of the first galaxies and large-scale structures led to a highly ionized intergalactic medium (IGM) below $z \lesssim 6$ \citep[for recent reviews see][]{Dayal2018,Wise2019,GnedinMadau2022,Robertson2022}.

The EoR represents a critical frontier in understanding cosmic evolution. The ionizing ultraviolet (UV) radiation from intense star-forming regions and active galactic nuclei (AGN) during this epoch ionized the host haloes and ambient IGM in proximity to these sources. Local environmental factors played a significant role in this process, particularly influencing star formation in the lowest-mass galaxies of the early Universe. Photoheating feedback could evaporate gas from the shallow potential wells of newly forming `minihaloes' with dark matter halo masses $M_\text{vir} \sim 10^6\,\Msun$ \citep{Dijkstra2004,Shapiro2004}. In contrast, `atomic cooling haloes' with masses above $M_\text{vir} \gtrsim 10^8\,\Msun$ and virial temperatures exceeding the $\sim10^4\,\text{K}$ threshold for cooling due to atomic hydrogen, were less affected by external radiation, experiencing only minor suppression of small-scale fragmentation and changes in cold flow accretion \citep{OhHaiman2002}.

The first galaxies formed during the EoR are the precursors of present-day galaxies and preserve the imprints of reionization in fossil remnants of old stellar populations in our Galactic backyard and local cosmic neighbourhood \citep{BrommYoshida2011}. However, the EoR presents significant observational challenges. The deepest \textit{Hubble Space Telescope} (\textit{HST}) observations have identified approximately a thousand galaxy candidates between redshifts $z=6$--$8$ and only a handful at higher redshifts \citep{Bouwens2015,Livermore2017,Atek2018,Bouwens2022}. The \textit{James Webb Space Telescope} (\textit{JWST}) is rapidly extending our observational reach. The advancements in high-$z$ galaxy surveys and spectroscopic follow-up are largely due to the unprecedented infrared wavelength coverage to access and constrain rest-UV and optical emission from high-redshift galaxies \citep[e.g.][]{Tacchella2023,Atek2023,Furtak2023,Fujimoto2023}. Moreover, probing the ionization and temperature structure of the IGM during reionization is also challenging. However, radio interferometers targeting the 21\,cm line of neutral hydrogen, such as the Murchison Widefield Array (MWA), Low-Frequency Array (LOFAR), Hydrogen Epoch of Reionization Array (HERA), and Square Kilometer Array (SKA), are providing a more complete picture of the state of the IGM \citep[e.g.][]{HERA2022}.

In our own cosmic neighborhood, dwarf galaxies in the Local Group are thought to resemble the most abundant galaxies during the EoR, particularly in terms of stellar mass and metallicity, though differences in other properties such as star formation rate or escape fraction may exist. Their low surface brightness complicates the detection of their high-$z$ progenitors \citep{Patej2015}. Observations suggest that star formation in some of these galaxies may have been suppressed due to reionization \citep{Weisz2014,Skillman2017,Brown2014,Bettinelli2018}, while others do not exhibit clear signatures of such suppression \citep{Weisz2014,Skillman2017,Grebel2004,Monelli2010,Hidalgo2011}. Connecting high- and low-redshift phenomena is of great interest, as it provides insights into the evolution and properties of galaxies and the IGM over cosmic time.

Numerical simulations incorporating ionizing radiation are essential for interpreting recent EoR observations in detail. Yet, modelling reionization is uniquely challenging due to its complex, multi-scale nature \citep{Ciardi2003,Iliev2006,Cain2023}. Accurate simulations require simultaneously capturing the large-scale processes of global reionization and the detailed physics of individual galaxy formation, including dark matter dynamics, gas physics, star and black hole formation, feedback mechanisms, and radiation hydrodynamics \citep[RHD;][]{Vogelsberger2020Review}. Additionally, an extraordinary number of resolution elements are necessary to accurately track atomic cooling haloes across a statistically significant volume of the Universe \citep{Gnedin2014,Pawlik2017,Rosdahl2018,Ocvirk2020,Kannan2022,Lewis2022,Katz2023}.

In this study, we employ the high-resolution \thesan suite of large-volume cosmological reionization simulations \citep{Kannan2022,Smith2022,Garaldi2022,Garaldi2024}. Leveraging the adaptive moving mesh magneto-hydrodynamics code \arepo \citep{Springel2010,Weinberger2020}, in combination with the state-of-the-art IllustrisTNG galaxy formation model \citep{Weinberger2017,Pillepich2018,Springel2018}, self-consistent RHD \citep{Kannan2019}, and dust modelling \citep{McKinnon2017}, our work utilises the \thesanone (RHD) and \thesandarkone (dark matter only) simulations to offer unique insights into galaxy and IGM properties at both high ($z \gtrsim 5.5$) and low ($z \sim 0$) redshifts. Through these sister simulations, we model the impact of reionization on galaxy formation in the Local Group and analyse the global reionization process, including the evolution of ionized bubbles in overdense environments.

Despite its success on large scales, the $\Lambda$CDM paradigm continues to face persistent tensions on small scales, most notably the `missing satellite problem' \citep{Klypin1999,Moore1999} and the `too-big-to-fail problem' \citep{Boylan-Kolchin2011}. Cosmological simulations predict an abundance of low-mass dark matter haloes, yet the observed number of luminous dwarf galaxies in the Local Group is significantly lower. A leading explanation is that cosmic reionization quenched star formation in low-mass haloes by suppressing gas accretion and cooling during their early evolution \citep{Bullock2017}. Observations of ultra-faint dwarfs in the Milky Way suggest that many of them formed the bulk of their stars before $z \sim 10$, consistent with early suppression of star formation due to reionization feedback \citep[e.g.][]{Brown2014}. UV feedback sterilizes the lowest-mass haloes by removing gas, leaving behind quenched stellar populations \citep{Benitez2015}. However, the contribution of low-mass galaxies to reionization remains debated. Some studies argue that a steep faint-end galaxy luminosity function was required to sustain reionization \citep[see][for discussion]{Bouwens2015,Finkelstein2015}, whereas others question whether these galaxies produced enough ionizing photons\citep[e.g.][]{Ferrara2013}.  

Because reionization was a highly inhomogeneous process, its impact on galaxy formation must be understood in a spatial and environmental context \citep{Iliev2006}. Denser regions, hosting clustered sources, tend to reionize earlier \citep{Li2014}, whereas underdense regions lag behind due to delayed structure formation and weaker emissivity \citep{Dixon2018}. This spatial variation leaves observable signatures. Many Local Group dwarf galaxies show early bursts of star formation that ceased around the EoR, consistent with reionization quenching \citep{Milosavljevic2014,Aparicio2016}, and recent work suggests that halo properties at $z = 0$ correlate with their reionization timing \citep{Zhu2019}.

Yet it remains unclear how reionization proceeded in group environments like the Local Group, and how halo interactions influenced the quenching of their satellites. The Local Group provides an ideal setting to address these questions, owing to its well-characterized satellite populations and dynamical structure. Recent studies have suggested that group-scale dynamics can play a non-negligible role in shaping reionization histories. For instance, \citet{Sorce2022} compared Local Group analogues to isolated halo pairs and found that factors such as halo mass, pair binding energy, and separation can modulate the timing of reionization. In this work, we explore how reionization timing correlates with local environmental conditions and how this connection shapes the evolution of galaxies in group environments.

This paper is structured as follows. In Section~\ref{sec:methods}, we describe the \thesan suite of cosmological RHD and dark matter-only simulations, and outline the methodology for tracking reionization histories of present-day haloes. Section~\ref{sec:environment} investigates the impact of large-scale density fluctuations on the timing of reionization. In Section~\ref{sec:local_group}, we detail the selection of Local Group analogues and characterize the reionization histories of their progenitors. Section~\ref{sec:virgo} examines the role of nearby Virgo-like clusters in modulating reionization timing. Finally, we conclude with a summary of our main findings and their implications in Section~\ref{sec:conclusions}. Supplementary discussions on the completeness of bijective subhalo matching are provided in Appendix~\ref{appendix:bijection}, on smoothing effects in Appendix~\ref{appendix:smoothing}, and on the definition of reionization redshift in Appendix~\ref{appendix:HII-threshold}.

\section{Methods}
\label{sec:methods}

In this section, we outline the simulations and data analysis techniques employed in our study. We utilise the \thesan simulations to link present-day dark matter properties with the reionization histories of Local Group analogues.

\subsection{THESAN simulations}

\begin{figure*}
  \includegraphics[width=\textwidth]{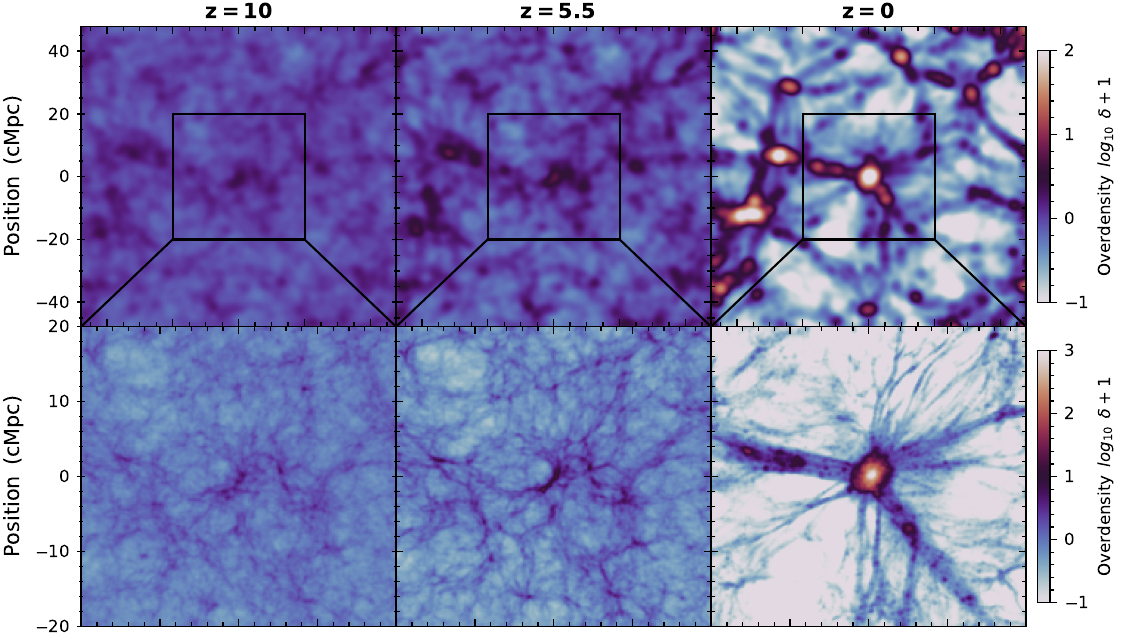}
  \caption{Visualization of the spatial distribution of dark matter overdensity $\delta = \rho/\overline{\rho} - 1$, shown as a 3 cMpc deep projection across the entire $(95.5\,\text{cMpc})^2$ simulation box (top panels) and a zoomed-in $(40\,\text{cMpc})^2$ region centered on the most massive halo of \thesandarkone (bottom panels), at redshifts $z = \{10, 5.5, 0\}$ from left to right. The top (bottom) panels apply Gaussian smoothing scales of $\sigma_\text{smooth} = 1\,\text{cMpc}~(125\,\text{ckpc})$. The initially overdense regions, which reionize early, continue to grow more overdense over time, while initially underdense regions, which reionize late, become increasingly underdense. The value of the overdensity is sensitive to the smoothing scale, which propagates into reionization properties.}
  \label{fig:overd_projection}
\end{figure*}

The \thesan simulations are a suite of large-volume, high-resolution cosmological RHD simulations that self-consistently model galaxy formation and reionization \citep{Kannan2022,Smith2022,Garaldi2022,Garaldi2024}. They employ the moving-mesh magneto-hydrodynamics code \areport \citep{Kannan2019}, an extension of the \arepo code \citep{Springel2010,Weinberger2020}, which incorporates radiative transfer (RT) capabilities. \areport solves the fluid equations on an unstructured Voronoi mesh that adapts to the fluid flow, enabling an accurate quasi-Lagrangian treatment of cosmological gas dynamics. The first two moments of the RT equation are solved assuming the M1 closure relation \citep{Levermore1984}, and the scheme reaches second-order accuracy.

The \thesan simulations use the state-of-the-art and well-tested IllustrisTNG galaxy formation model \citep{Weinberger2017,Pillepich2018}, an updated version of the original Illustris simulation framework \citep{Vogelsberger2013, Vogelsberger2014b, Vogelsberger2014a}, including detailed treatments of star formation, feedback processes, and black hole physics. Dust physics is also incorporated following \citet{McKinnon2017}. Gravity is calculated using a hybrid Tree-PM approach: long-range forces are computed using a particle-mesh algorithm, while short-range forces are calculated using a hierarchical oct-tree method. To minimize force errors during time integration, node centres are randomly shifted at each domain decomposition, as detailed in \citet{Springel2021}. The simulations also employ a hierarchical time integration scheme to efficiently handle the wide range of dynamical timescales involved.

The RT equations are coupled to a non-equilibrium thermochemistry solver to accurately compute the ionization states of hydrogen and helium, as well as temperature changes due to photo-ionization, metal cooling, and Compton cooling. A reduced speed of light approximation (RSLA) is used for computational efficiency, with an effective speed of light of $\tilde{c} = 0.2\,c$ \citep[see Appendix~A of][]{Kannan2022}. For practicality, only the hydrogen ionizing Lyman continuum (LyC) part of the radiation spectrum is modelled. Photons are discretized into energy bins defined by the thresholds: [13.6, 24.6, 54.4, $\infty$)\,eV. Each cell in the simulation tracks the comoving photon number density and flux for each bin. To compensate for the limited frequency resolution, the radiation in each bin is assumed to follow the spectrum of a 2\,Myr old, quarter-solar metallicity stellar population, computed using the Binary Population and Spectral Synthesis models \citep[BPASS v2.2.1;][]{Eldridge2017,Stanway2018}, assuming a Chabrier IMF \citep{Chabrier2003}. The amplitude is set by the photon number density in the cell.

The \thesan simulations have a box size of $L_\text{box} = 95.5\,\text{cMpc}$, assuming cosmological parameters from \citet{Planck2018} with $h = 0.6774$, $\Omega_m = 0.3089$, $\Omega_\Lambda = 0.6911$, and $\Omega_b = 0.0486$. In this work, we primarily utilise the flagship high-resolution simulation, \thesanone, along with its dark matter-only counterpart, \thesandarkone. The simulation contains $2100^3$ dark matter and (initial) gas particles for resolutions of $m_\text{DM} = 3.12 \times 10^6\,\Msun$ and $m_\text{gas} = 5.82 \times 10^5\,\Msun$, respectively, effectively resolving atomic cooling haloes by at least $50$ particles. While \thesanone runs from high redshift down to $z = 5.5$, \thesandarkone extends all the way to $z = 0$, enabling us to trace the evolution of dark matter structures to the present day. By combining these two simulations, we can study the reionization histories of Local Group analogues in a self-consistent manner.
This complements other studies from the \thesan simulations, including emission line intensity mapping \citep{Kannan2022b}, the impact of reionization on low-mass galaxies \citep{Borrow2023}, constraining galaxy populations \citep{Kannan2023}, ionizing escape fractions \citep{Yeh2023}, Lyman-$\alpha$ emitters \citep{Xu2023}, alternative dark matter models \citep{Shen2024a}, galaxy sizes \citep{Shen2024b}, ionized bubble statistics \citep{Neyer2024, Jamieson2025}, and galaxy--IGM connections \citep{Garaldi2024a, Garaldi2024b, Kakiichi2025}, as well as \textsc{thesan-zoom} for more accurately capturing the multi-phase interstellar medium \citep{Kannan2025}.

\begin{figure*}
	\includegraphics[width=\textwidth]{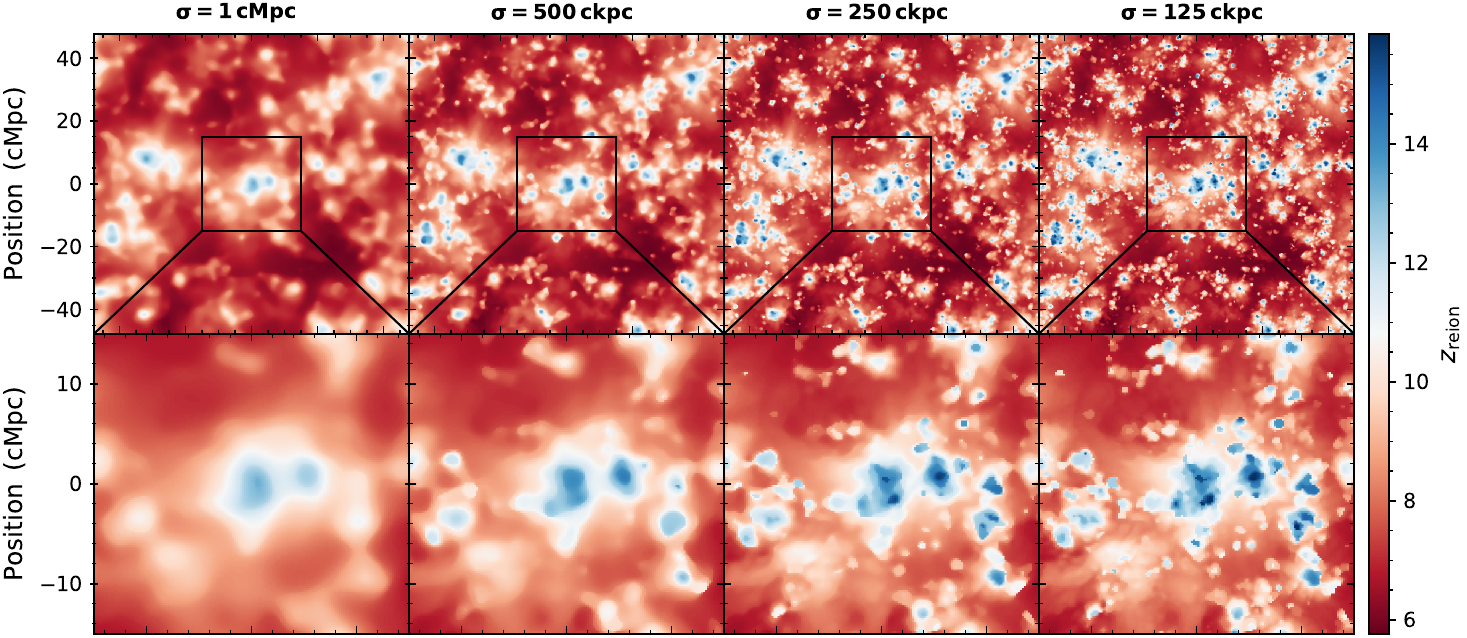}
	\caption{Visualization of the spatial distribution of reionization redshift $z_\text{reion}$ centred on the most massive halo of \thesandarkone, using Gaussian smoothing scales of $\sigma_\text{smooth} = \{\text{1\,cMpc, 500\,ckpc, 250\,ckpc, 125\,ckpc}\}$ from left to right. The reionization redshift is defined as the latest time when the volume-smoothed neutral hydrogen fraction drops below the threshold $x_\text{HI} = 0.5$. The lower panels show that small-scale structures in $z_\text{reion}$ are increasingly suppressed for larger smoothing scales, due to the infilling of small-scale ionized bubbles. Regions that reionized earliest correspond to the highest overdensity regions in Fig.~1, and the large-scale correlation persists despite smoothing-induced delays in $z_\text{reion}$.}
	\label{fig:zreion_slice}
 \end{figure*}

\subsection{Catalogue construction}
We now describe the data products utilised in our analysis. Our primary focus is connecting present-day ($z = 0$) properties of dark matter haloes to their reionization histories at $z \gtrsim 5.5$. To facilitate this, we construct catalogues of large-scale environmental properties for all 80 snapshots of the \thesan simulations, spanning redshifts from $z = 20$ to $z = 5.5$ \citep[also used by][]{Neyer2024}. First, we bin the particles onto $1024^3$ Cartesian grids and compute the dark matter overdensity, $\delta \equiv (\rho - \bar{\rho}) / \bar{\rho}$, where $\rho$ is the local density and $\bar{\rho}$ is the mean density. To study the environmental dependence at different scales, we apply a periodic, mass-conserving Gaussian smoothing kernel defined by standard deviations of 0\,ckpc (no smoothing), 125\,ckpc, 250\,ckpc, 500\,ckpc, and 1\,cMpc. We also store additional quantities such as the volume-weighted ionized fraction. For analyses requiring spatial information, we use rendered grids coarse-grained to 512 cells per side, while halo catalogues are inherited from the positions within the $1024^3$ grids.

To accurately determine environmental properties of Local Group analogues, we utilise the merger trees from \thesandarkone to trace haloes from $z = 0$ back to $z = 5.5$. We then employ cross-links between \thesandarkone and \thesanone at $z = 5.5$ to identify corresponding haloes in \thesanone, and further trace their progenitors to higher redshifts using the merger trees from \thesanone.

For deeper explorations into the formation and evolution of cosmic structures, we construct environmental merger tree catalogues by tracking the main progenitors of all subhaloes in \thesanone back to their formation redshifts, $z_\text{form}$. From the histories of each group and subhalo, we calculate the redshift of reionization, $z_\text{reion}$, defined as the latest time when the volume-smoothed neutral fraction drops below a given threshold. Our fiducial value is $x_\text{\HI} = 0.5$, but we also explore thresholds of $0.1$ and $0.9$ to assess the impact of different reionization definitions (see Appendix~\ref{appendix:HII-threshold}). Since $z_\text{reion}$ is an interpolated value, we also record the local overdensity and subhalo mass at these epochs. To assign $z_\text{reion}$ to the $z = 0$ sample, we create an atlas of bijective cross-linked subhaloes, i.e. ones that map in both directions between \thesanone and \thesandarkone, which ensures a maximally complete and reliable sample for $z=0$ analyses. This means a $z_\text{reion}$ can be assigned to a $z=0$ \thesandarkone subhalo if its DM-only merger tree traces back to a bijective match to \thesanone at $z=5.5$. We quantify the completeness of this bijective matching in Appendix~\ref{appendix:bijection}, and find that it is nearly perfect for subhaloes above $\sim 10^9\,\Msun$ at $z = 5.5$.

We have also produced catalogues containing the environmental properties of groups and subhaloes for all snapshots of \thesanone and \thesandarkone simulations, along with their medium resolution \thesantwo and \thesandarktwo counterparts. Dark matter haloes are identified by the Friends-of-Friends (FoF) algorithm \citep{Davis1985}, and subhaloes are identified using the SUBFIND algorithm \citep{Springel2001}. For each snapshot, we generate HDF5 files containing datasets for groups (FoF haloes) and subhaloes, including properties such as temperature, \HII fraction, overdensity, and the \HI photoionization rate $\Gamma_\text{\HI}$. These properties are also computed using the same smoothing scales from 125\,ckpc to 1\,cMpc. For each smoothing scale, we also generate a summary grid file containing the redshift of reionization at three different ionized fraction thresholds, $x_\text{\HI} = \{0.1, 0.5, 0.9\}$, as well as the formation redshift of the main progenitor and subhalo mass at $z_\text{reion}$ when $x_\text{\HI} = 0.5$, self-consistently tracking the time-dependent locations and environments of the haloes. This allows us to investigate the effects of smoothing scale and reionization threshold on our results (see Appendix~\ref{appendix:HII-threshold}).

For the DM-only simulation \thesandarkone, we cannot directly measure baryonic environmental properties. Therefore, we infer these properties by mapping the positions of haloes in \thesandarkone to the corresponding positions in \thesanone. This positional correspondence allows us to estimate quantities like temperature, ionization fraction, and overdensity for haloes in \thesandarkone, which is crucial for studies requiring baryonic information but are based on DM-only simulations.
By combining the strengths of the DM-only simulation \thesandarkone and the baryonic simulation \thesanone, we can perform comprehensive studies of the properties and evolution of subhaloes. This approach allows us to select Local Group candidates using \thesandarkone and incorporate detailed baryonic physics from \thesanone \citep[with minimal gas biases for the MW-like targets compared to smaller satellites; see][]{Vogelsberger2014a}, providing a continuous understanding of cosmic structures at high- and low-redshifts.

\section{Environmental properties}
\label{sec:environment}

\begin{figure*}
	\includegraphics[width=\textwidth]{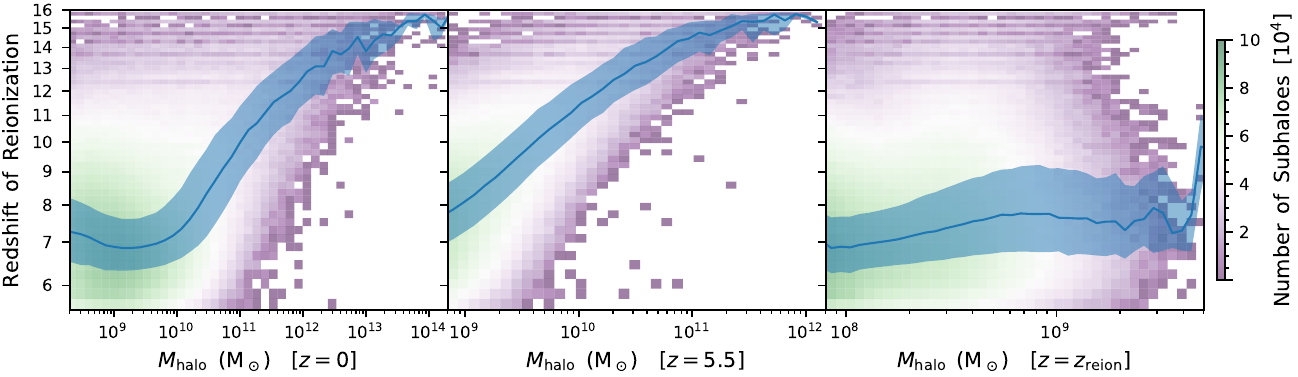}
    \caption{Correlation between reionization redshift $z_\text{reion}$ and halo mass $M_\text{halo}$ at $z = 0$ (left panel), $z = 5.5$ (middle panel), and at the individual reionization redshift of each halo $z = z_\text{reion}$ (right panel), using $\sigma_\text{smooth} = 125\,\text{ckpc}$. The solid lines show the median values, and the shaded regions indicate the $16^\text{th}$ to $84^\text{th}$ percentiles. The background colour scale represents the halo number density in each two-dimensional bin. At $z = 0$, a positive correlation persists for haloes with $M_\text{halo} \gtrsim 10^{10}\,\Msun$, although the overall trend is weakened by nonlinear growth. At $z = 5.5$, a strong positive correlation is evident across the full mass range, particularly for the most massive haloes. At $z = z_\text{reion}$, the correlation is not yet fully developed, as many low-mass haloes are externally reionized at early times, while high-mass haloes span a broad range in $z_\text{reion}$, reflecting a combination of inside-out self-ionization and delayed exposure to external radiation fields.}
    \label{fig:m_zreion}
\end{figure*}

\begin{figure*}
	\includegraphics[width=\textwidth]{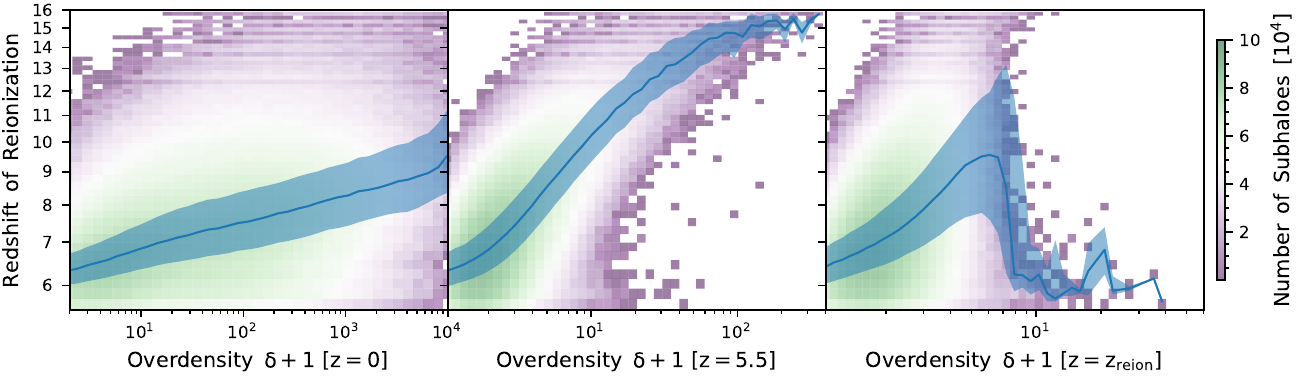}
    \caption{Correlation between reionization redshift $z_\text{reion}$ and environmental overdensity $\delta$ at $z = 0$ (left panel), $z = 5.5$ (right panel), and $z = z_\text{reion}$ (right panel), using $\sigma_\text{smooth} = 125\,\text{ckpc}$. The lines show the medians, and the shaded regions indicate the $16^\text{th}$ to $84^\text{th}$ percentiles. The colour bar shows the distribution density of haloes. Haloes in higher overdensity regions tend to reionize earlier, although the variation due to small-scale structure is important as well. The trend is better preserved down to $z = 0$ as overdensities at $z = 5.5$ include proto-cluster progenitors. The peak around $\delta \sim 2\!-\!5$ in the right panel reflects inside-out reionization. The smooth decline in $z_\text{reion}$ toward lower overdensities may reflect delayed or flash reionization in more diffuse environments, where external radiation dominates and local sources are limited.}
\label{fig:overd_zreion}
\end{figure*}

In this section, we explore the relationships between environmental properties such as overdensity, reionization redshift, and halo mass, and how they influence structure formation and the reionization process. We utilise the smoothed spatial grid data, applying a Gaussian filter with a standard deviation of $\sigma_\text{smooth} = 125\,\text{ckpc}$. This smoothing scale is close to the Jeans or filtering scale \citep{Gnedin1998,Gnedin2000}, which is relevant for understanding the suppression of gas accretion in low-mass haloes due to photoheating feedback.

\subsection{Spatial distribution of overdensity}
In Fig.~\ref{fig:overd_projection}, we present visualizations of the spatial distribution of the volume-weighted dark matter overdensity $\delta = \rho / \overline{\rho} - 1$ within full-box (upper panels; $95.5 \times 95.5 \times 3\,\text{cMpc}^3$) and zoomed-in (lower panels; $40 \times 40 \times 3\,\text{cMpc}^3$) subvolumes centred on the most massive halo of \thesandarkone, at redshifts of $z = \{10, 5.5, 0\}$ from left to right. The overdensity in the upper (lower) panels are smoothed on scales of $\sigma_\text{smooth} = 1\,\text{cMpc}~(125\,\text{ckpc})$ to show the range of filters used in this study. The highest overdensity regions correspond to the locations of the most massive haloes, where the earliest galaxies form due to the dense environments facilitating accelerated gravitational collapse and star formation.

At early times $z \gtrsim 10$, the Universe is in the quasi-linear regime of structure formation, and densities are relatively smooth, only beginning to show significant structures. The small fluctuations in density ($\delta \sim 10^{-5}$ at $z \approx 1100$ imprinted in the cosmic microwave background) eventually grow under gravity into highly nonlinear structures, most prominent on smaller scales and later times. The cosmic web emerges with massive haloes experiencing rapid growth from gas accretion and mergers, leading to the formation of galaxies. By $z = 0$, the Universe exhibits more complex structures. Dense regions have evolved into well-defined clusters and filaments, while underdense regions have become voids. The evolution from the nearly homogeneous state at high redshift to the structured state at low redshift reflects the hierarchical nature of cosmic structure formation \citep{Maulbetsch2007, Ocvirk2020}.

In general, initially overdense regions, which reionize early, continue to become more overdense, while initially underdense regions, which reionize late, become more underdense, as demonstrated in several previous studies \citep{Dawoodbhoy2018,Kannan2022}. The value of the overdensity is sensitive to the smoothing scale of the Gaussian filter, which propagates to various reionization properties as well. We discuss the impact of different smoothing scales more in Appendix~\ref{appendix:smoothing}.

\begin{figure}
    \includegraphics[width=\columnwidth]{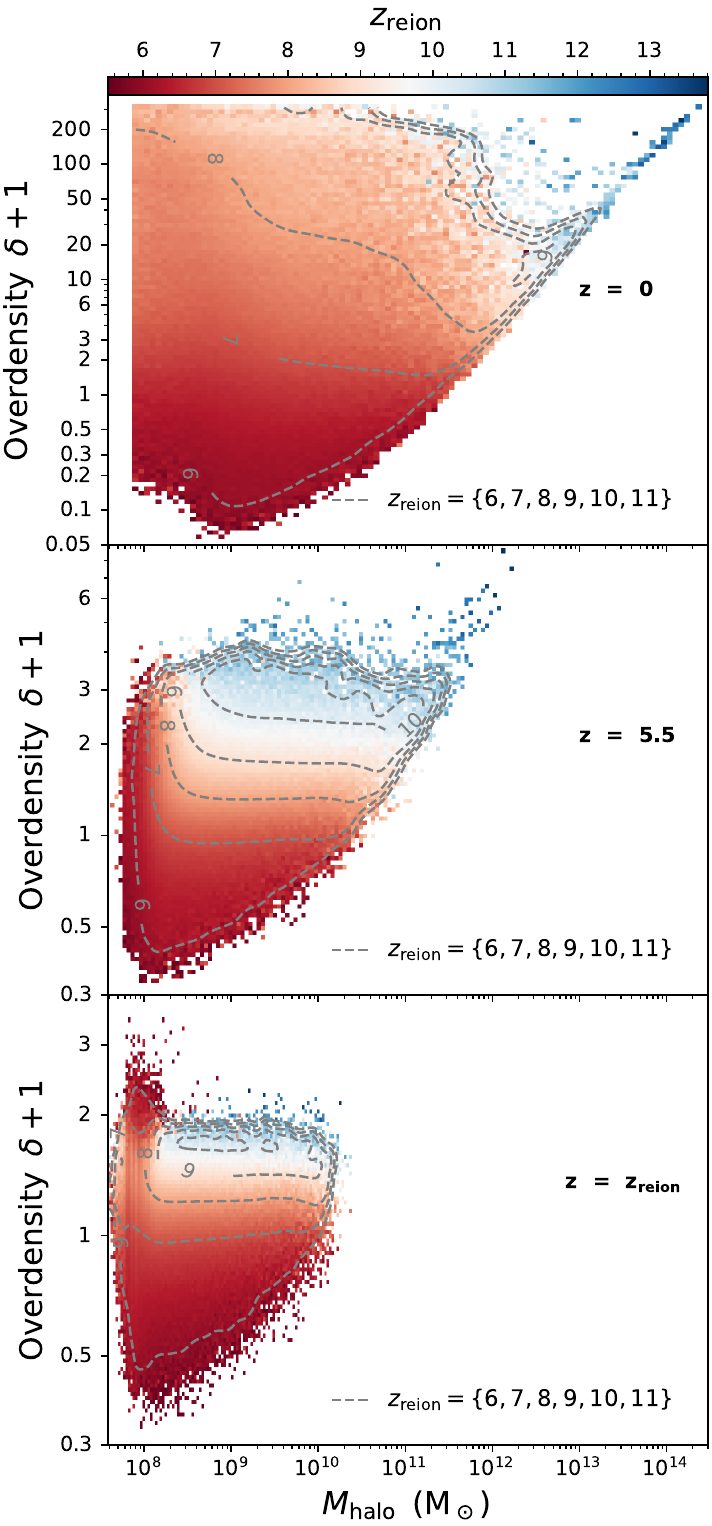}
    \caption{Two-dimensional distribution of halo mass $M_\text{halo}$ and overdensity $\delta$, using a smoothing scale of $\sigma_\text{smooth} = 1\,\text{cMpc}$, shown at $z = 0$ (top panel), $z = 5.5$ (middle panel), and the individual reionization redshift of each halo $z = z_\text{reion}$ (bottom panel). The colour scale indicates the average reionization redshift in each bin. Dashed contours trace integer levels of the levels of the average $z_\text{reion}$ across the 2D plane. More massive haloes located in overdense regions typically reionize earlier, reflecting the combined influence of halo mass and local environment. An increasing overdensity threshold with halo mass traces the centers of inside-out reionization, while late-reionizing haloes are primarily located in underdense environments or at the low-mass end.}
    \label{fig:hist_overd_m}
\end{figure}

\subsection{Spatial distribution of reionization redshift}
We compare the spatial distribution of overdensity with that of the reionization redshift $z_\text{reion}$, defined as the redshift at which a given region remains persistently ionized, to examine how initial density fluctuations shape the timing and pattern of reionization. In Fig.~\ref{fig:zreion_slice}, we present visualizations of these $z_\text{reion}$ spatial distributions centred on the most massive halo, for different smoothing scales $\sigma_\text{smooth} = \{\text{1\,cMpc, 500\,ckpc, 250\,ckpc, 125\,ckpc}\}$. The reionization redshifts are also sensitive to the choice of smoothing scale because local and environmental \HII fractions become less correlated with larger filtering volumes. Large smoothing scales disperse the \HI density distribution, causing infilling of small bubbles, making the ionization maps appear smoother, and delaying the reionization redshifts. Still, the inhomogeneity of reionization and its large-scale correlations are evident in these maps. Regions that reionized earliest generally correspond to the highest overdensity regions observed in Fig.~\ref{fig:overd_projection}. These areas likely hosted the very first galaxies and intense star-forming activity, contributing significantly to the early reionization process. Conversely, regions with lower reionization redshifts indicate areas that remained neutral longer, often corresponding to underdense or void regions beyond the reach of significant early star formation.

The gradient of reionization redshifts reflects the hierarchical nature of structure formation, where dense regions formed early structures that generated the LyC radiation necessary for reionization, while underdense regions lagged behind. The connection between the evolving spatial distributions of overdensity and \HII fraction illustrates the critical role of initial density fluctuations in shaping the timeline of cosmic reionization. Overall, there is a strong positive correlation between overdensity and reionization redshift, with the exact details being sensitive to the smoothing scale, highlighting the patchiness of reionization.

\begin{figure}
	\includegraphics[width=\columnwidth]{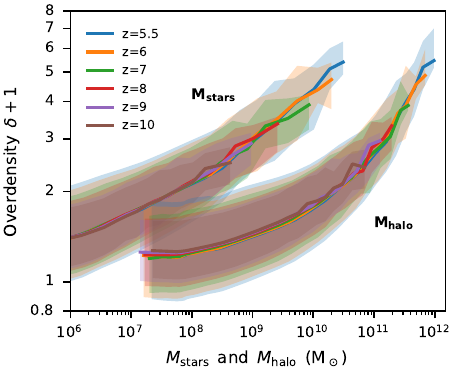}
    \caption{Median overdensity as a function of stellar and halo mass for haloes in the \thesanone simulation at redshifts $z = \{5.5, 6, 7, 8, 9, 10\}$. Solid lines indicate the medians, and shaded regions represent the $16^\text{th}$ to $84^\text{th}$ percentile ranges. A clear positive correlation is seen across all redshifts, with substantial environmental scatter but minimal evolution in the overall trend. The stellar-to-halo mass relation follows the expected scaling, with the scatter reflecting variations in local environment and formation history.}
    \label{fig:T1_overd_m}
\end{figure}

\begin{figure*}
	\includegraphics[width=\textwidth]{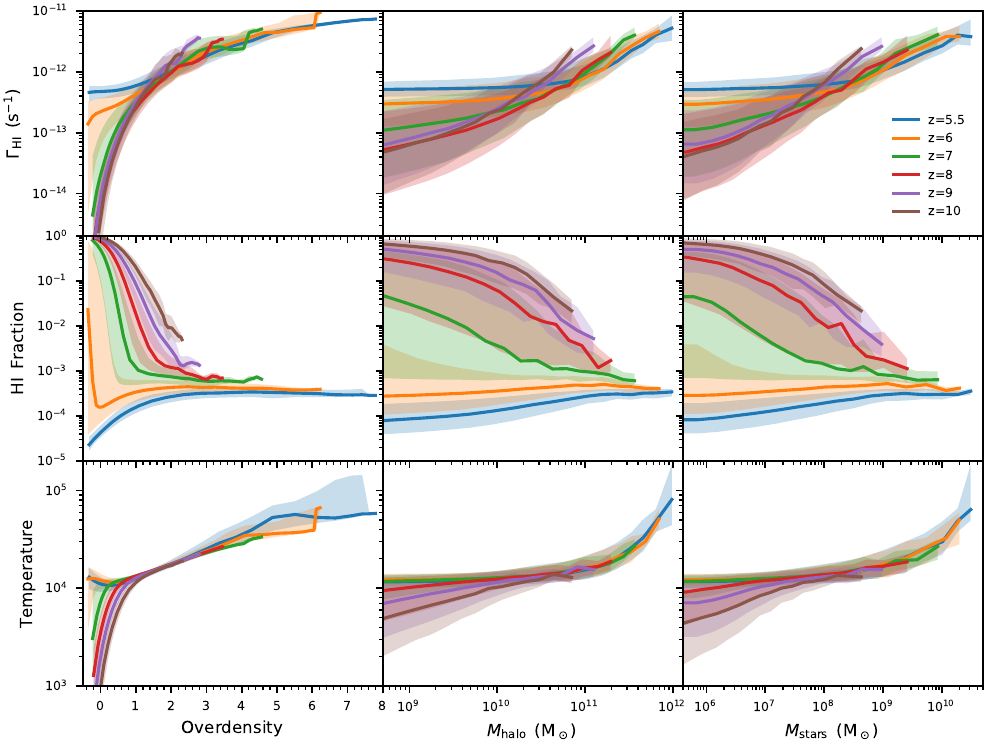}
    \caption{Volume-weighted median values of gas temperature $T$, \HI fraction, and photoionization rate $\Gamma_{\HI}$ as functions of overdensity (left column), halo mass (middle column), and stellar mass (right column), for haloes with $M_\text{halo} > 10^8\,\Msun$ in the \thesanone simulation. Results are shown at redshifts $z = \{5.5, 6, 7, 8, 9, 10\}$. Solid lines indicate the medians, and the shaded regions represent the $16^\text{th}$ to $84^\text{th}$ percentile ranges. Densest and most massive haloes show the highest $\Gamma_{\HI}$ and temperatures, along with the lowest neutral fractions, reflecting their earlier reionization and stronger radiative feedback. In contrast, low-mass and underdense environments exhibit delayed reionization and cooler, more neutral gas.}
    \label{fig:T1_properties}
\end{figure*}

\subsection{Correlation between reionization redshift and halo mass}
We investigate the statistical correlation between halo mass $M_\text{halo}$ and reionization redshift $z_\text{reion}$ using the subset of haloes with well-resolved progenitor histories extending beyond $z = 5.5$. We utilise the merger tree catalogues to trace the progenitors of haloes at $z = 0$ back to $z = 5.5$ and employ the cross-links between \thesandarkone and \thesanone simulations to assign $z_\text{reion}$ values to haloes. We adopt subhalo masses at $z = 0$ from \thesandarkone and at $z = 5.5$ and $z_\text{reion}$ from \thesanone. A halo is assigned a reionization redshift only if it has a progenitor at $z > 5.5$. We find that a smoothing scale of $\sigma_\text{smooth} = 125\,\text{ckpc}$ yields the strongest correlation, so in the remainder of the paper we adopt this value. The impact of this choice is explored in Appendix~\ref{appendix:smoothing}, and the effect of the threshold ionized fraction is discussed in Appendix~\ref{appendix:HII-threshold}.

In Fig.~\ref{fig:m_zreion}, we illustrate the relationship between $z_\text{reion}$ and $M_\text{halo}$ at three critical epochs: the present-day ($z = 0$; left panel), the end of the reionization era ($z = 5.5$; middle panel), and at the specific reionization redshift of each subhalo ($z = z_\text{reion}$; right panel). The lines represent the median values, and the shaded regions indicate the $16^\text{th}$ to $84^\text{th}$ percentiles for the asymmetric one sigma variation. The background colours show the full distributions of all haloes above the resolution limit of the simulation, i.e. $M_\text{halo} \gtrsim 10^8 \Msun$.

At $z = 0$ (left panel), haloes with masses $M_\text{halo}(z=0) \gtrsim 10^{10}\,\Msun$ tend to have reionized earlier than lower-mass ones, indicating that the progenitors of massive haloes were located in denser regions with higher densities of ionizing sources. However, it appears that the imprint of the reionization history is only partially preserved, as the initial correlation between $z_\text{reion}$ and halo mass becomes less pronounced due to the significantly more complex large-scale structure evolution from $z = 5.5$ to $z = 0$. The scatter in $z_\text{reion}$ among low-mass haloes is particularly large, reflecting the delayed exposure to ionizing radiation for haloes residing in less-connected or underdense regions \citep[e.g.,][]{Alvarez2009}.

At $z = 5.5$ (middle panel), a strong positive correlation between $z_\text{reion}$ and $M_\text{halo}$ is evident, with nearly all of the most massive haloes (e.g., $M_\text{halo} \gtrsim 10^{12}\,\Msun$) reionizing at early times, reflecting their early formation in highly overdense regions. In contrast, low- and intermediate-mass haloes populate a wider portion of the $z_\text{reion}$ distribution, ranging from $z_\text{reion} \sim 5.5$ to $z_\text{reion} \gtrsim 9$, indicating that their reionization histories reflect a broad range of local environmental conditions, as their large abundance allows more complete sampling of such variation. For low-mass haloes, $M_\text{halo} (z = 0) \lesssim 10^{10}\,\Msun$, the tight correlation observed at $z = 5.5$ is largely erased by $z = 0$, as lower-mass haloes may overtake initially larger haloes in mass through hierarchical growth at later times ($z < 5.5$).

Finally, at the individual $z = z_\text{reion}$ of each halo (right panel), the correlation with halo mass has not yet fully emerged. While there is a mild tendency for lower-mass haloes to reionize later, the overall trend remains weak. This behavior, together with the relatively narrow mass range at this epoch, is partly driven by the abundance of low-mass haloes that are rapidly reionized by external ionizing fronts (see the green regions of the distribution). Many of these haloes reionize at relatively high redshifts, despite their low masses, as they are located near early-forming sources of ionizing radiation. At the same time, inside-out reionization allows haloes to self-ionize once they grow to a sufficient mass, without relying on external ionizing sources. This critical threshold roughly corresponds to the atomic cooling halo mass, $\sim 10^9\,\Msun$ at $z \sim 6$, above which haloes can sustain star formation and reionize their surroundings independently.

\subsection{Correlation between reionization redshift and overdensity}
Regions that reionized earliest are generally associated with high overdensities, motivating a more detailed examination of how $z_\text{reion}$ correlates with local density as traced by haloes—complementing the volume-averaged results in \citet{Neyer2024}. In Fig.~\ref{fig:overd_zreion}, we show this relation at $z = 0$ (left panel), $z = 5.5$ (middle panel), and at the individual reionization redshift $z = z_\text{reion}$ of each halo (right panel), adopting $\sigma_\text{smooth} = 125\,\text{ckpc}$.

At $z = 5.5$, $z_\text{reion}$ increases steeply and smoothly with overdensity across $\delta \sim 10$--$100$, reflecting the earlier ionization of haloes in dense regions due to clustered sources and enhanced radiation fields. The positive correlation remains largely intact by $z = 0$, reflecting the coherent evolution of large-scale density structures, though it becomes weaker, likely due to nonlinear evolution between the end of reionization and the present day. Because we follow halo merger trees, the evolution of both $z_\text{reion}$ and overdensity is properly tracked over time. While a clear correlation is maintained, local deviations arise due to small-scale structure, stochastic assembly histories, and non-local ionizing sources. Overall, these trends are consistent with previous findings that overdense regions reionize early, while voids reionize later \citep{Dawoodbhoy2018}.

The right panel of Fig.~\ref{fig:overd_zreion}, evaluated at $z = z_\text{reion}$, exhibits a broad peak around $\delta \sim 2$--$5$, consistent with inside-out reionization driven by clustered sources in moderately overdense regions. The smooth decline toward lower overdensities is consistent with delayed ionization in more diffuse environments, possibly involving haloes reionized by external radiation, with limited local sources or partial self-shielding.

\begin{figure*}
  \centering
  \includegraphics[width=\textwidth]{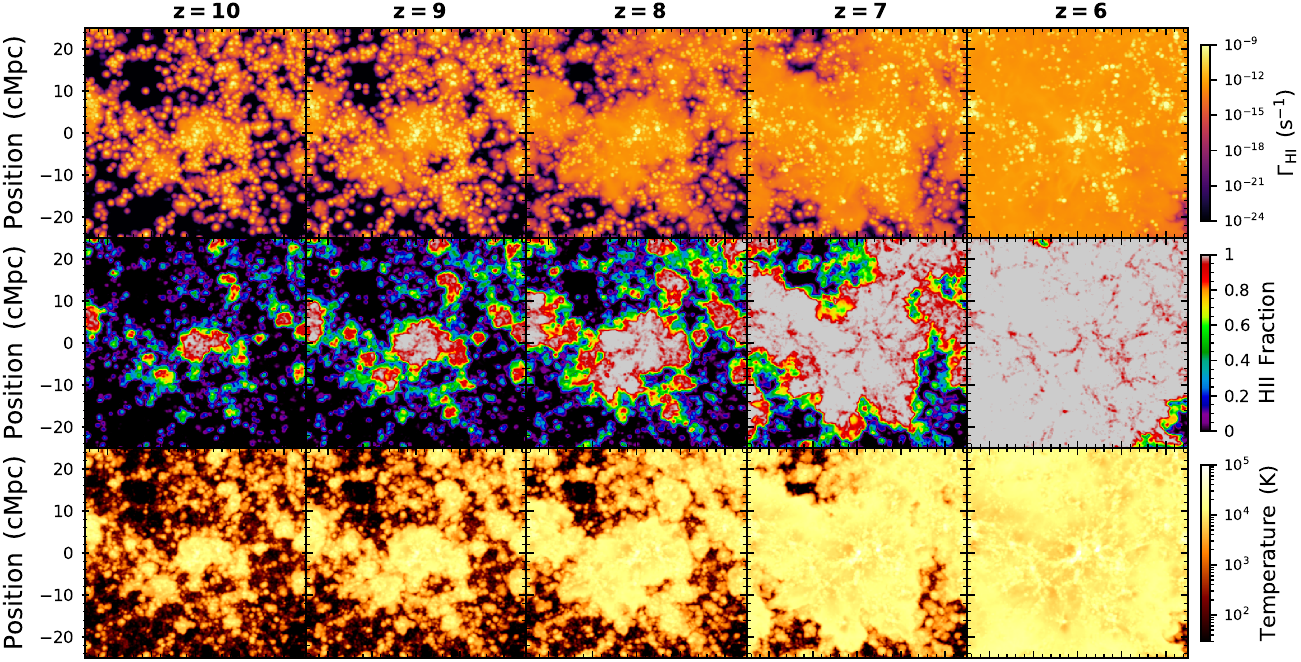}
  \caption{Projection over a 3 cMpc depth showing the evolution of photoionization rate $\Gamma_\text{HI}$ (top), \HII fraction (middle), and  gas temperature (bottom) around the most massive halo in \thesandarkone, by tracing its progenitor positions from $z = 10$ to $z = 6$ using the cross\_link algorithm, and extracting ionization-related fields from the fully coupled \thesanone simulation. The maps show volume-weighted, unsmoothed quantities within a $50~\text{cMpc}$ region. Ionized regions expand outward from a compact core at high redshift into an extended zone by $z = 6$, with $\Gamma_\text{HI}$, \HII fraction, and temperature closely aligned near the halo centre. The delayed ionization in outer regions reflects an extreme inside-out reionization process.}
  \label{fig:projection}
\end{figure*}

We present in Fig.~\ref{fig:hist_overd_m} two-dimensional histograms of halo mass $M_\text{halo}$ versus overdensity $\delta$ at $z = 0$ (top panel), $z = 5.5$ (middle panel), and $z = z_\text{reion}$ (bottom panel), with the colour bar indicating the average $z_\text{reion}$, and dashed contours marking integer values of this average, again using $\sigma_\text{smooth} = 125\,\text{ckpc}$. These distributions illustrate a richer perspective of how both halo mass and overdensity influence the timing of reionization, showing stronger gradients in $\delta$ at both $z = 5.5$ and $z = z_\text{reion}$. The marginally resolved halo range of $M_\text{halo} \lesssim 2\times10^8\,\Msun$ is dominated by the significant population of low-mass haloes with little star-formation, late assembly, or external late reionization. 

However, more massive haloes show a significant environmental dependence in the range $z_\text{reion} \in [7, 11]$. We also observe that the minimum overdensity required to host a halo increases with halo mass. This rising floor reflects the confinement of massive haloes to denser environments, consistent with the expectation that reionization proceeds inside-out, beginning in the highest-density regions. In fact, haloes more massive than $M_\text{halo} \gtrsim 10^{11}\,\Msun$ at $z = 5.5$ (or $M_\text{halo} \gtrsim 10^{12}\,\Msun$ at $z = 0$) exhibit a strong connection with overdensity that drives earlier $z_\text{reion}$ values for nearby lower-mass haloes as well. By $z = 0$, most late-reionizing haloes remain outside overdense regions ($\delta \lesssim 2$). However, a sufficient number of low-mass haloes fall into overdense regions without substantial growth that the $z_\text{reion} \sim 8$ contour is shaped equally by both mass and overdensity. Overall, the structure and $z_\text{reion}$ gradients across different regions in the $M_\text{halo}$--$\delta$ plane demonstrate the importance of both local and environmental factors during the EoR.

\begin{figure}
	\includegraphics[width=\columnwidth]{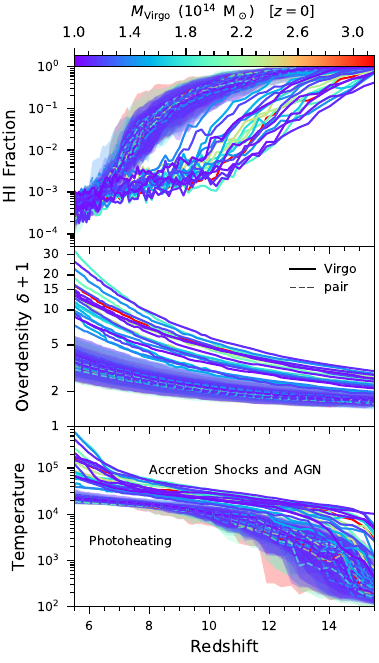}
	\caption{Evolution of volume-weighted \HI fraction (top), overdensity $\delta + 1$ (middle), and gas temperature (bottom) around Virgo-like haloes, smoothed at $\sigma_\text{smooth} = 500\,\text{ckpc}$. Solid lines show the median profiles of Virgo-like haloes, while dashed lines represent nearby Milky Way/Andromeda-like haloes with ($M_\text{halo}(z=0) \in [8 \times 10^{11}, 3 \times 10^{12}]\,\Msun$). Shaded regions indicate the $16^\text{th}$–$84^\text{th}$ percentile range for the Milky Way/Andromeda-like haloes. Virgo haloes experience an early and rapid drop in \HI fraction between $z \sim 16$ and $z \sim 11$, along with an early temperature rise, consistent with reionization driven by clustered sources in overdense regions. In contrast, LG analogues show a delayed and more gradual evolution, with both reionization and heating progressing slowly due to their lower-density environments.}
	\label{fig:LG_properties}
\end{figure}

\begin{figure}
	\includegraphics[width=\columnwidth]{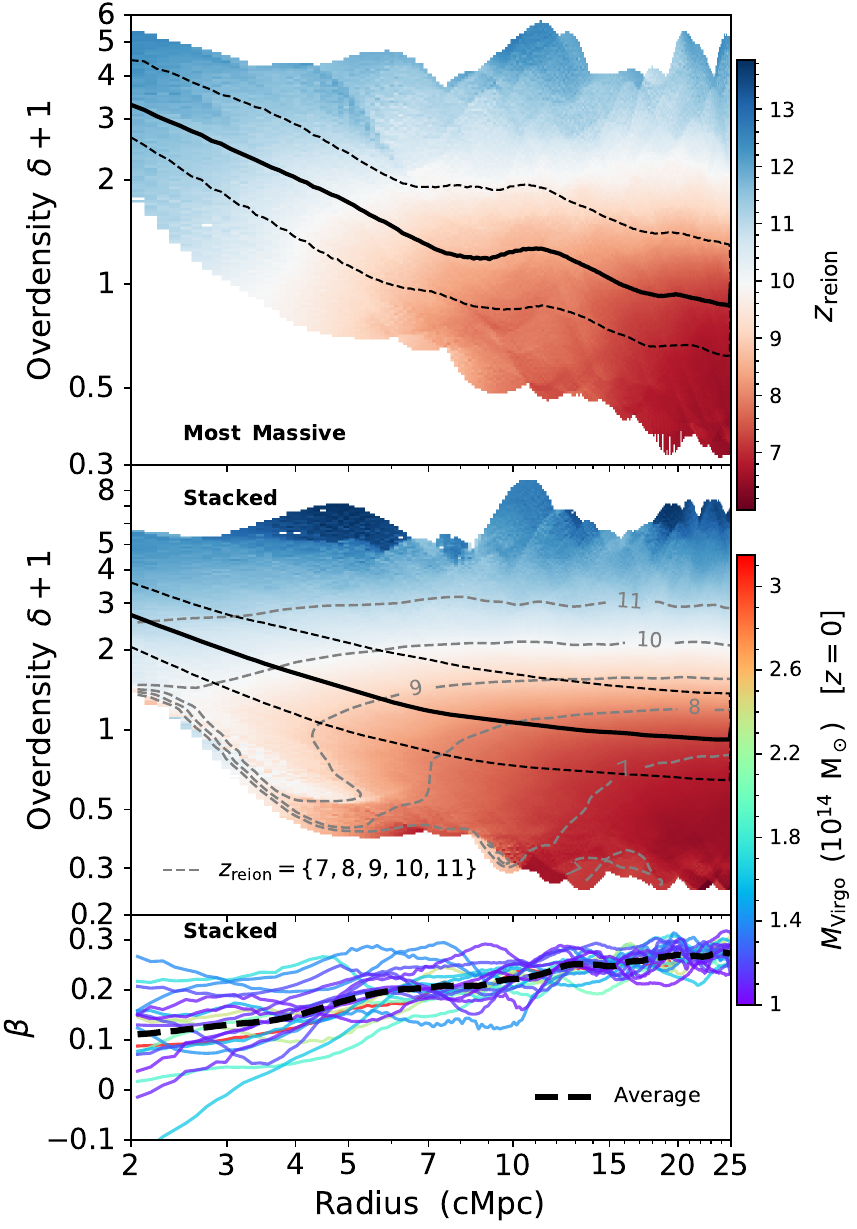}
	\caption{Two-dimensional histograms showing how the reionization redshift $z_\text{reion}$ depends on overdensity and radial distance from Virgo-like haloes, with all quantities smoothed at $\sigma_\text{smooth} = 1\,\text{cMpc}$. The top panel presents the most massive Virgo-like halo, while the middle panel shows the stacked distribution for all 20 Virgo analogues. The stacked panel includes contours of constant $z_\text{reion}$, which show its variation with overdensity and radius. Regions closer to halo centers reionize earlier, and $z_\text{reion}$ increases with overdensity at fixed radius. The bottom panel shows the radial profiles of the logarithmic slope $\beta = \text{d}\log z_\text{reion} / \text{d}\log(\delta + 1)$, quantifying the strength of the $z_\text{reion}$–density correlation. The increase of $\beta$ with radius reflects a transition from central source-dominated to environment-driven reionization, consistent with an extreme inside-out scenario.}
	\label{fig:hist}
\end{figure}

\begin{figure*}
  \includegraphics[width=\textwidth]{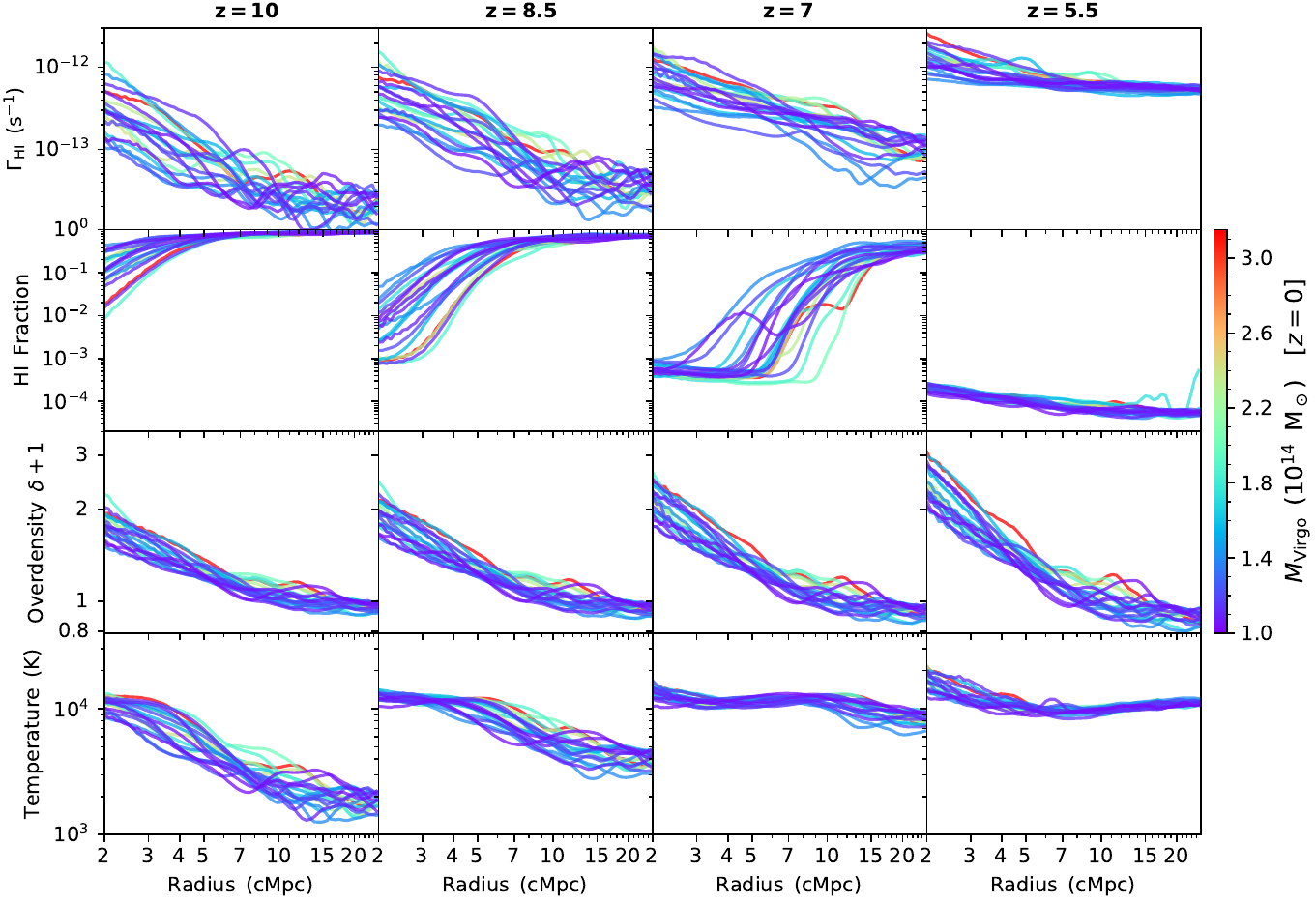}
  \caption{Spherically averaged radial profiles of photoionization rate $\Gamma_{\rm HI}$ (top row), \HI fraction (second row), dark matter overdensity $\delta + 1$ (third row), and gas temperature (bottom row) around Virgo-like haloes at $z = \{5.5, 7, 8.5, 10\}$. All quantities are volume-weighted and smoothed at $\sigma_\text{smooth} = 500\,\text{ckpc}$ to emphasize large-scale trends beyond $\sim 1\,\text{cMpc}$. Line colors indicate the $z = 0$ halo mass of each Virgo analogue. Ionization and heating proceed from the inside out, with early, localized enhancements in $\Gamma_{\rm HI}$ and temperature near halo centers that gradually extend outward. Overdensity profiles remain stable in shape but increase in amplitude over time, while temperature evolution slightly lags behind ionization, reflecting the propagation of heating fronts.}
  \label{fig:radial_profiles}
\end{figure*}

\begin{figure}
	\includegraphics[width=\columnwidth]{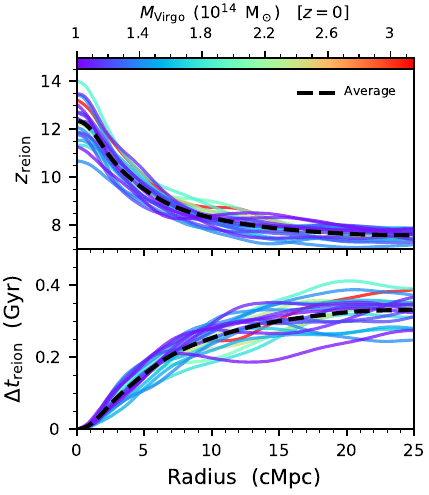}
	\caption{Radial profiles of reionization timing around Virgo-like haloes, smoothed at $\sigma_\text{smooth} = 1\,\text{cMpc}$. The top panel shows $z_\text{reion}$, which declines from $z \sim 12$ near halo centers to $z \sim 9$ at $r \sim 10\,\text{cMpc}$, consistent with inside-out reionization driven by early-forming, overdense regions. The bottom panel shows the relative quantities $\Delta t_\text{reion} = t_\text{reion}^\text{Virgo} - t_\text{reion}(r)$, which quantify the reionization timing offset between the central halo and its surroundings. Both exhibit steep gradients within $r \lesssim 4\,\text{cMpc}$, beyond which the delay grows more gradually. At $r \gtrsim 8\,\text{cMpc}$, reionization occurs up to 200–300 Myr later than at the halo center, indicating a transition to background-dominated reionization at distances where the central halo no longer governs the timing. Thick dashed black curves show the average profiles across all Virgo analogues.}
	\label{fig:profile_z_reion}
\end{figure}

\subsection{Physical properties of haloes at different redshifts}
The strong correlations between overdensity and halo mass highlight the spatial organization of ionized regions around early-forming structures. Fig.~\ref{fig:T1_overd_m} shows the average overdensity as a function of halo mass and stellar mass at redshifts $z = \{5.5, 6, 7, 8, 9, 10\}$. Both relations exhibit similar trends, with overdensity increasing monotonically with mass. Halo masses are typically $\sim 100\times$ larger than the stellar component, with the stellar-to-halo mass ratio ($M_\text{stars} / M_\text{halo}$) declining toward lower-mass haloes, consistent with reduced star formation efficiency in less massive systems \citep{Duffy2010}.

Fig.~\ref{fig:T1_properties} presents the \HI photoionization rate $\Gamma_{\HI}$ (top row), \HI fraction (middle row), and gas temperature $T$ (bottom row) as functions of overdensity (left column), halo mass (middle), and stellar mass (right), using $\sigma_\text{smooth} = 125\,\text{ckpc}$. All quantities are shown as volume-weighted means within resolved haloes with $M_\text{halo} > 10^8\,\Msun$. Multiple redshifts are shown to track the evolution across reionization, specifically $z = \{5.5, 6, 7, 8, 9, 10\}$.

For lower-mass haloes, these properties are sensitive to redshift, reflecting ongoing reionization and heating of the IGM. In higher-mass haloes, the properties show less variation with redshift, as these haloes reionized earlier (i.e. $z_\text{reion} \gtrsim 10$) and their environments have stabilized, as shown in the middle row. Specifically, in terms of the neutral fraction, the redshift dependence is significantly reduced at $\delta \gtrsim 3$, $M_\text{halo} \gtrsim 10^{11}\,\Msun$, and $M_\text{stars} \gtrsim 10^{9}\,\Msun$. The greatest variation is near the midpoint of reionization around $z \sim 7$, as the conditions are more isolated at early times and more homogeneous at late times. Overall, there are no significant qualitative differences between halo and stellar mass dependencies.

The \HI photoionization rate, $\Gamma_{\HI}$, increases with all overdensity and halo mass, due to the higher abundance of ionizing sources in dense clustered regions. This is especially the case during the first half of reionization, but remains true at the higher $\delta$, $M_\text{halo}$, and $M_\text{stars}$ end, even after the UV background $\Gamma_{\HI}$ has begun to set in. The \HI fraction decreases with increasing overdensity and halo mass, most dramatically from $7 < z < 10$. The temperature increases with overdensity and halo mass, reflecting stronger heating and more intense feedback processes influencing the surrounding gas. The relationship between temperature and overdensity is similar to results from the late reionization models at $z = 4$ in \citet{Trac2008}. However, the imprint of a rapid, earlier evolution is also observed as these trends highlight the interplay between structure formation, reionization, and the thermal evolution of the IGM.

\section{Local group properties}
\label{sec:local_group}

The Local Group (LG) of galaxies is part of the Local Supercluster, a flattened distribution of galaxies centered on the Virgo cluster. It consists of the Milky Way, the Andromeda Galaxy (M31), and at least 80 other members—mostly dwarf galaxies—with the two dominant galaxies each having halo masses of approximately $10^{12}\,\Msun$ \citep[e.g.][]{McConnachie2012,Boylan-Kolchin2013}. The system exhibits a characteristic ``dumbbell'' shape: the Milky Way and its satellites form one lobe, while Andromeda and its satellites form the other. The two lobes are separated by about 800 kpc and are moving toward each other at a relative velocity of 123 km/s \citep{vanderMarel2012}. The LG is embedded in an intergalactic medium of hot gas and dark matter, which affects gas accretion and galaxy interactions.

\begin{figure*}
\includegraphics[width=\textwidth]{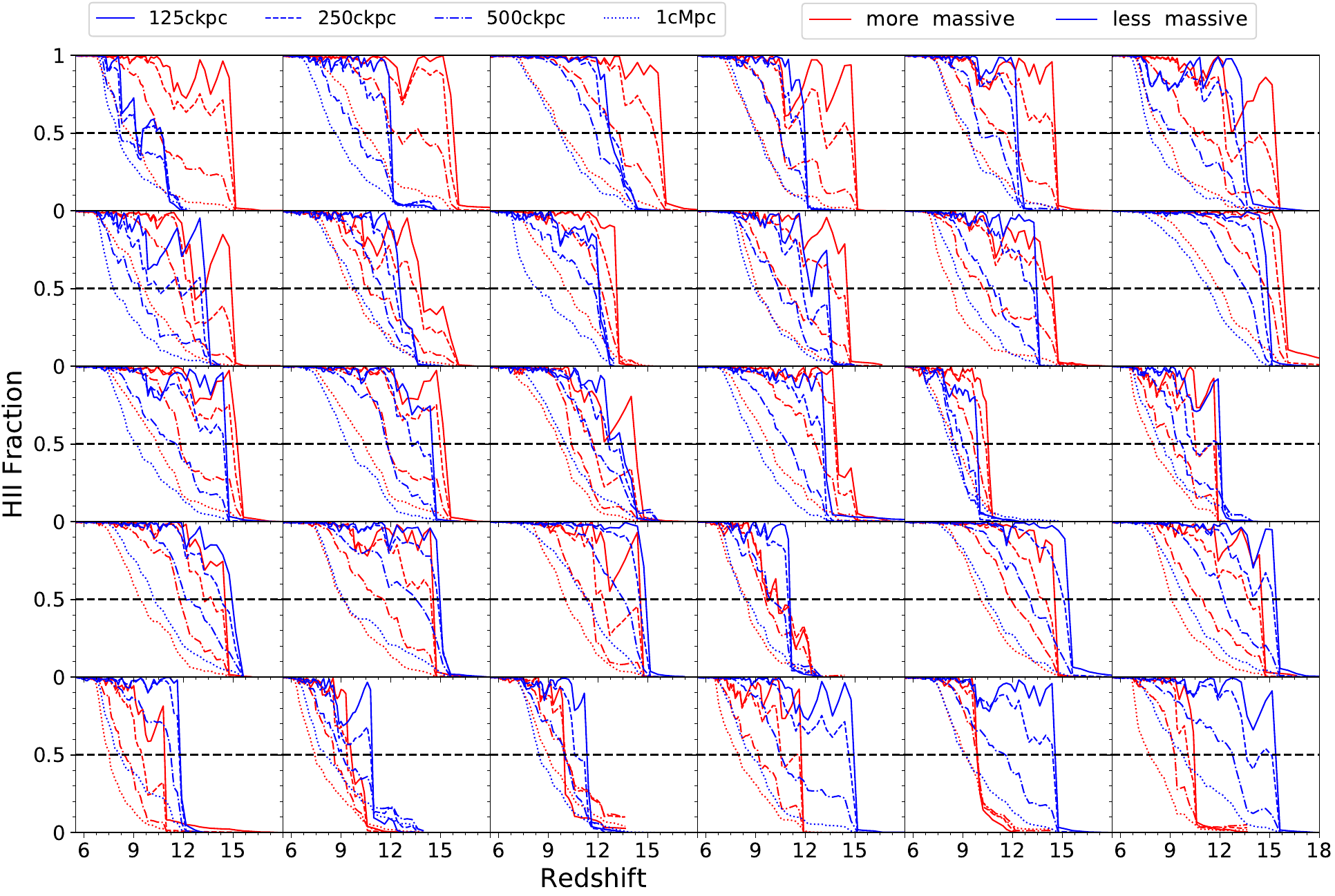}
	\caption{Evolution of \HII fraction for a subset of Local Group analogues. Each panel shows one LG pair, with the more massive (M31-like) halo plotted in red and the less massive (MW-like) halo in blue. The curves are derived from \thesanone, tracing progenitors identified via the cross\_link algorithm applied to \thesandarkone at $z = 5.5$. Reionization redshifts are computed from the merger trees in \thesanone. Different line styles indicate the smoothing scale used to define $z_\text{reion}$: $\sigma_\text{smooth} = \left\{125\,\text{ckpc}, 250\,\text{ckpc}, 500\,\text{ckpc}, 1\,\text{cMpc}\right\}$. At small smoothing scales, the reionization histories appear more irregular, reflecting local environmental variations, while larger scales yield smoother profiles and reduce the apparent timing offset. Examples can be found where either the more or less massive halo reionizes earlier, with reionization redshift differences of up to $\sim 1$--$2$ in some cases.}
 \label{fig:HII_history}
\end{figure*}

\subsection{Selection of Local Group analogues}
We select 20 Virgo-like haloes from \thesandarkone, for which the halo mass is above $10^{14}\,\Msun$ at $z = 0$. Following a similar LG pair selection criteria adopted by previous studies \citep[e.g.,][]{Garrison-Kimmel2014,Libeskind2020,Sorce2022}, we identify pairs based on the following conditions:
\begin{itemize}
  \item Subhalo Mass: $8 \times 10^{11} < M_\text{halo}/\Msun < 3 \times 10^{12}$ 
  \item Separation: $0.5 < d_\text{sep} / \text{cMpc} < 2.5$
  \item Isolation: no third halo is more massive than the smaller one within $2\,\text{cMpc}$ of the midpoint position
  \item Mass Ratio: the smaller-to-larger halo mass ratio must be $> 0.5$
  \item Proximity to a Virgo Analogue: the pair must be located within $20\,\text{cMpc}$ from a Virgo analogue, i.e. $M_\text{Virgo} > 10^{14}\,\Msun$
\end{itemize}
We do not restrict the pairs to be approaching each other ($v_\text{rad} < 0$), or exclude pairs within $20\,\text{cMpc}$ of more than one Virgo analogue. Following these selection criteria, we identified 224 LG pairs that resemble the Milky Way--Andromeda system within the influence of a Virgo-like cluster. For reference, the sample size would be 3840 pairs if the proximity criterion is relaxed, i.e. they are not near a Virgo-like cluster but satisfy the other criteria.

\subsection{Impact of reionization on Local Group pairs}
\label{sec:impactofrei}

\begin{figure}
    \centering
    \includegraphics[width=\linewidth]{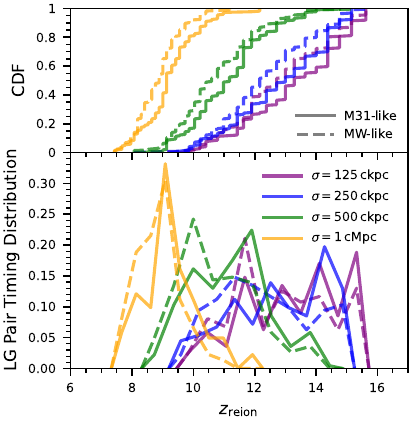}
    \caption{Distribution of reionization redshifts ($z_\text{reion}$) for haloes in Local Group pairs, separated into the more massive (M31-like; solid) and less massive (MW-like; dashed) haloes. The top panel shows the cumulative distribution function (CDF), while the bottom panel displays the corresponding probability density. Curves are shown for smoothing scales of $\sigma_\text{smooth} = \left\{\text{125\,ckpc, 250\,ckpc, 500\,ckpc, 1\,cMpc}\right\}$. For M31-like haloes, the median $z_\text{reion}$ values with $1\sigma$ confidence intervals are $\{13.28^{+1.86}_{-1.85}, 12.99^{+1.38}_{-2.01}, 11.18^{+1.20}_{-1.31}, 9.13^{+1.09}_{-0.83}\}$, respectively. For MW-like haloes, the values are $\{12.71^{+2.04}_{-1.68}, 12.16^{+1.84}_{-1.52}, 10.59^{+1.30}_{-1.17}, 8.88^{+0.66}_{-0.70}\}$. The difference between the two halo populations is most evident at smaller smoothing scales, where $z_\text{reion}$ distributions are broader and more distinct. At larger smoothing scales, the environments are more spatially uniform, resulting in later, more similar $z_\text{reion}$ distributions for the two haloes.}
    \label{fig:zreion_hist}
\end{figure}

\begin{figure}
\includegraphics[width=\columnwidth]{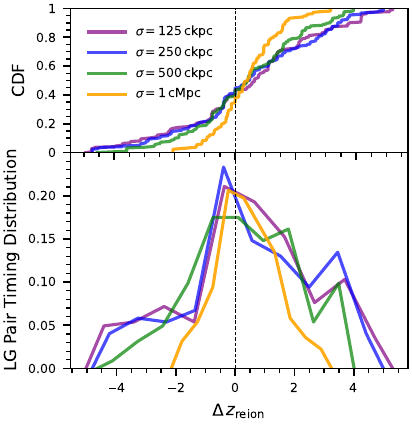}
	\caption{Distribution of reionization redshift differences between haloes in Local Group pairs, $\Delta z_\text{reion} = z_\text{reion}^1 - z_\text{reion}^2$, where $z_\text{reion}^1$ and $z_\text{reion}^2$ are the more and less massive haloes, respectively. The top (bottom) panel shows the cumulative (probability) distribution functions at smoothing scales of $\sigma_\text{smooth} = \left\{\text{125\,ckpc, 250\,ckpc, 500\,ckpc, 1\,cMpc}\right\}$. Median values and $1\sigma$ confidence regions of $\Delta z_\text{reion}$ are $\left\{{0.64^{+2.34}_{-2.79},\ 0.38^{+2.61}_{-2.13},\ 0.43^{+1.78}_{-1.78},\ 0.41^{+1.11}_{-1.07}}\right\}$, respectively. At all smoothing scales, the more massive halo in each pair tends to reionize earlier on average. However, significant scatter exists, especially at smaller smoothing scales, where $\Delta z_\text{reion}$ can be negative, indicating instances in which the less massive halo reionized first.}
 \label{fig:dt_histogram}
\end{figure}

Fig.~\ref{fig:projection} shows the evolution of the local ionization environment surrounding the most massive halo in the \thesandarkone simulation at $z = 0$, by tracing its progenitor positions at $z = \left\{6, 7, 8, 9, 10\right\}$ within the fully coupled \thesanone box using the cross\_link algorithm. The figure shows volume-weighted, unsmoothed maps of the photoionization rate (top), ionized hydrogen fraction (middle), and gas temperature (bottom) within a $50~\text{cMpc}$ region. These maps capture the temporal evolution of ionized regions surrounding the most massive halo, progressing from a compact ionized core at $z = 10$ to an extended ionized zone by $z = 6$. The spatial alignment of $\Gamma_{\text{HI}}$, \HI fraction, and gas temperature near the halo centre, together with delayed ionization in the outer regions, is indicative of an inside-out reionization mode. In this scenario, ionized bubbles originate around early-forming sources embedded in overdense peaks and gradually expand into the surrounding intergalactic medium, as also reproduced in simulations such as CoDa~II \citep{Ocvirk2020}.

\begin{figure*}
  \centering
  \includegraphics[width=\textwidth]{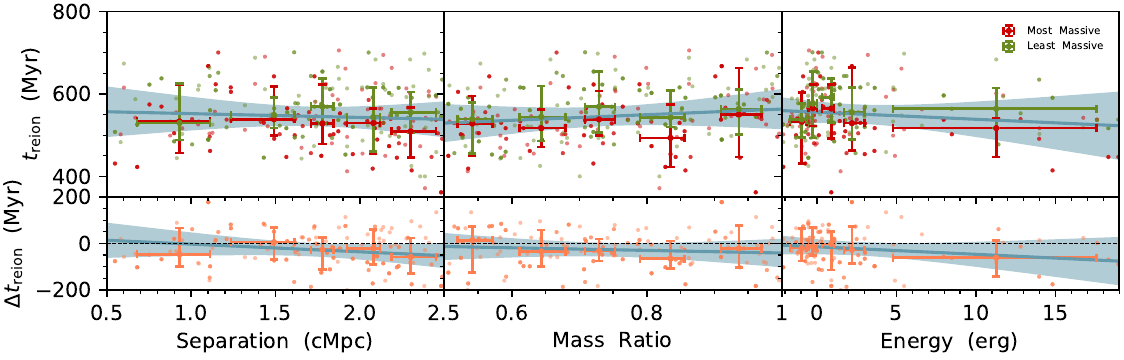}
  \caption{Top row: reionization times of the more massive (red) and less massive (green) haloes in each Local Group pair, plotted against present-day pair separation (left), mass ratio (middle), and total energy (right). Bottom row: reionization time difference between the two haloes, $\Delta t_\text{reion}$, shown as a function of the same quantities. For each panel, the pairs are sorted by the corresponding $x$-axis quantity and divided into five equal-size bins (44 haloes each); the error bars show the median and interquartile range within each bin. Median $t_\text{reion}$ values diverge with increasing separation; $\Delta t_\text{reion}$ is largest in widely separated or intermediate mass ratio pairs, and shows no clear dependence on total energy.}
  \label{fig:time_sepa_mt_E}
\end{figure*}

Fig.~\ref{fig:LG_properties} compares the reionization histories of Virgo-like haloes and LG analogues in terms of \HI Fraction, overdensity, and gas temperature. Virgo haloes experience an early and rapid drop in \HI between $z \sim 16$ and $z \sim 11$, driven by efficient photon production in overdense regions. In contrast, LG pairs follow a more gradual trajectory, with the steepest decline occurring between $z \sim 8$ and $z \sim 7$. This behaviour is expected for haloes residing in lower-density environments, where delayed structure formation and reduced photon flux postpone the onset of reionization \citep{Dixon2018}.

The thermal histories of the two populations further illustrate the role of environment. Virgo-like haloes exhibit an early and pronounced temperature rise, coinciding with the onset of reionization and likely driven by photoheating in overdense regions. While AGN are unlikely to dominate the global ionizing photon budget \citep{Hassan2018, Eide2020, Yeh2023}, their feedback may locally enhance ionization and heating, particularly in dense environments \citep[e.g.,][]{Finlator2009, BoschRamon2018}. In contrast, LG pairs present a more gradual temperature rise over an extended redshift interval, indicative of a reionization process primarily driven by local photoionization. This thermal evolution is consistent with expectations for low-density environments, where the absence of strong gravitational collapse suppresses shock heating, and where ionization proceeds more slowly due to the lower abundance and clustering of ionizing sources \citep[e.g.,][]{Daloisio2019}.

\subsection{Influence of Virgo-like clusters on reionization}
\label{sec:virgo}

Fig.~\ref{fig:hist} shows how $z_{\text{reion}}$ depends on both overdensity and distance from Virgo-like haloes. In the top panel, we focus on the most massive halo; the middle panel stacks all 20 Virgo analogues. All quantities in Fig.~\ref{fig:hist} are smoothed on scales of $1\,\text{cMpc}$. In both cases, regions near the halo center have higher $z_{\text{reion}}$, indicating earlier reionization, with $z_{\text{reion}}$ decreasing outward. $z_{\text{reion}}$ increases with overdensity at each radius, and also rises as one approaches the central halo at fixed density. These results show that both overdensity and proximity to massive haloes contribute to the spatial variation in $z_{\text{reion}}$.

To quantify the relative contributions of central sources and local environment, we compute the logarithmic slope $\beta \equiv d\log z_{\text{reion}} / d\log(\delta + 1)$, shown in the bottom panel of Fig.~\ref{fig:hist}. We find that $\beta$ increases with radius, suggesting a gradual shift from source-dominated to environment-driven reionization. In the inner regions ($r \lesssim 5$ cMpc), $z_{\text{reion}}$ is primarily controlled by the ionizing output of the central Virgo-like halo.

At larger distances, this influence weakens, and the reionization timing becomes increasingly sensitive to the local overdensity and the contributions of more diffuse sources. This spatial transition is again consistent with the inside-out reionization picture, where clustered sources ionize nearby overdense regions first and later extend into lower-density surroundings \citep{Iliev2006, Zahn2007}. A similar trend is seen in the CoDa~II simulation \citep{Ocvirk2020}, where massive haloes reionize their surroundings out to several several cMpc, in agreement with the radial extent we find in our Virgo-like systems. This radial dependence of $\beta$ reflects the changing balance between source-driven and environment-driven reionization, consistent with previous analyses of reionization structure and patchiness \citep{Daloisio2015, Park2016}.
The observed gradient in $\beta$ suggests that haloes near massive clusters were reionized early by external sources, potentially suppressing star formation in nearby low-mass satellites \citep[e.g.,][]{Efstathiou1992, Gnedin2000, Benitez2015}. This may help explain the old stellar populations seen in some Local Group dwarf spheroidals \citep{Weisz2014, Benitez2015}.

Fig.~\ref{fig:radial_profiles} presents the evolution of spherically averaged radial profiles of photoionization rate $\Gamma_{\rm HI}$ (top row), the neutral hydrogen fraction \HI (second row), the dark matter overdensity ($\delta + 1$, third row), and the temperature (bottom row), centered on Virgo-like haloes. These profiles are smoothed over $\sigma_{\rm smooth} = 500\,\text{ckpc}$ to focus on the large-scale environment beyond $\sim 1\,\text{cMpc}$, shown at $z = \left\{10, 8.5, 7, 5.5\right\}$, with line colour indicating the $z = 0$ mass of the corresponding Virgo-like halo.

At high redshift ($z = 10$), $\Gamma_{\rm HI}$ remains low at large radii, but increases significantly toward $r \sim 2$–3 cMpc, where the densest early sources reside. This radial rise reflects the initial confinement of ionizing photons to the halo cores, with the radiation field still strongly shaped by the local source distribution. At these early times, the mean free path---the characteristic distance an ionizing photon can travel before being absorbed---is limited by the abundance of neutral hydrogen and remains well below the horizon scale \citep{Furlanetto2004}. As a result, the ionizing background is highly sensitive to the spatial distribution of sources, and large fluctuations in $\Gamma_{\rm HI}$ naturally arise in environments where both emissivity and absorption are modulated by local overdensities \citep{Davies2016}.

As reionization progresses, $\Gamma_{\rm HI}$ rises across all radii, and the profiles flatten, particularly beyond $r \gtrsim 8$ cMpc, signaling the transition toward a volume-filling ionized IGM. By $z = 5.5$, the photoionization rate reaches $\sim 10^{-12}\,\text{s}^{-1}$ at intermediate radii ($r \sim 4$–8 cMpc), while maintaining a relatively uniform level at larger distances. Notably, a mild enhancement persists at these intermediate scales, with more massive haloes exhibiting slightly elevated $\Gamma_{\rm HI}$ values relative to their lower-mass counterparts. This suggests that residual fluctuations in the ionizing background may survive even after global overlap, reflecting the continued influence of source clustering and locally extended mean free paths in overdense regions \citep{Davies2016}. Although small in amplitude, such mass-dependent differences indicate that the post-reionization radiation field remains locally modulated by large-scale structure.

The second row of Fig.~\ref{fig:radial_profiles} shows the evolution of the radial profile of the \HI fraction around Virgo-like haloes. At $z = 10$, the IGM remains largely neutral across the full radial range, except for a sharp drop in \HI within $r \sim 2$–3 cMpc near the halo centers, indicating the presence of compact ionized regions formed by the earliest clustered sources. As reionization progresses, this transition zone systematically moves outward, reflecting the steady expansion of ionized regions into the surrounding medium. The \HI floor within the bubble declines gradually with decreasing redshift, and by $z = 5.5$ the \HI fraction falls below $10^{-4}$ at all radii shown, consistent with the completion of reionization across the volume. The overall profile evolution demonstrates the outward propagation of ionization fronts and the growing coherence of the ionized IGM as small ionized bubbles percolate and merge with larger ones on cosmological scales \citep{Trac2008, Iliev2014, Jamieson2025}.

The dark matter overdensity profiles exhibit a stable radial structure across redshifts, increasing steadily toward the centres of Virgo-like haloes. At all epochs, overdensities rise steeply toward the halo centre, while the outer profiles remain close to the cosmic mean, reflecting the underlying matter clustering produced by hierarchical structure formation. Although the shape remains similar with time, the amplitude increases from $z = 10$ to $z = 5.5$, consistent with continued nonlinear growth and the deepening of gravitational potential wells. The correlation between high overdensity, enhanced $\Gamma_{\rm HI}$, and elevated temperatures at small radii reflects the role of massive haloes in regulating the state of the surrounding IGM across several cMpc.

The temperature structure of the IGM co-evolves with the ionization state, albeit with a slight lag. At early times ($z = 10$), heating is confined to the inner $\sim$1--2\,cMpc, where local ionizing sources have begun to photoheat the surrounding gas. As radiation propagates outward the temperature profiles flatten to $\sim 10^4\,\text{K}$, with slightly delayed heating fronts relative to ionization fronts \citep{Trac2008, Daloisio2019}. Even at $z = 5.5$, the central regions remain hotter due to stronger photoionization and galaxy clustering. The modest residual variations at large radii reflect the finite response timescale and suggest that the IGM retains a memory of reionization timing and source clustering \citep{Keating2018}. Eventually, the radial structure begins to resemble a temperature--density relation shaped by photoheating and adiabatic cooling, as expected from early and recent models \citep{Hui1997, Wells2024}.

Fig.~\ref{fig:profile_z_reion} shows the radial dependence of reionization timing around Virgo-like haloes, based on the smoothed profiles of $z_\text{reion}$ and $\Delta t_\text{reion}$, evaluated at $\sigma_\text{smooth} = 1\,\text{cMpc}$. In the top panel, the $z_\text{reion}$ profile exhibits a clear decline from $z \sim 12$ at small radii ($r \lesssim 2\,\text{cMpc}$) to $z \sim 9$ at $r \sim 10\,\text{cMpc}$, highlighting that early-forming, overdense regions host the first ionizing sources and reionize ahead of their surroundings. The bottom panel shows the relative quantities $\Delta t_\text{reion} = t_\text{reion}^{\text{Virgo}} - t_\text{reion}(r)$, which more directly quantify the timing difference between the Virgo-like haloes and their surrounding environment. The $\Delta t_\text{reion}$ profile exhibits a pronounced gradient at small radii ($r \lesssim 3$-4 cMpc), beyond which the decline becomes more gradual. This trend reflects the limited spatial domain over which a massive halo and its immediate environment can causally influence the timing of reionization. 

Beyond $r \sim 5$–6\,cMpc, reionization is increasingly governed by local sources and the ambient ionizing background, rather than by the central halo. At $r \gtrsim 8$\,cMpc, reionization occurs with a delay of 200--300\,Myr relative to the Virgo-like halo center, comparable to the full duration of the EoR. This substantial offset marks the outer limit of the halo's influence and indicates a transition to a regime where reionization is governed primarily by local sources and the ambient ionizing background. This universal feature can be summarized as the dominance of early biased structures giving way to spatially uncorrelated, patchy reionization fronts \citep{Zahn2007}.

\subsection{Differential timing in reionization in Local Group pairs}
\label{sec:reionizationtime}
In Fig.~\ref{fig:HII_history}, we show the reionization histories of a subset of LG pairs sorted by $\Delta z_\text{reion}$, which is also a halo mass-correlated timeline. On average, the more massive halo (M31 analog; red curves) reionized earlier compared to the less massive halo (MW analog; blue curves). However, in our biased sorted sample there are examples of the opposite behavior. While this seems counterintuitive, LG pairs were not always close to each other and could have been assembled in different environments or had different mass evolution histories. Both progenitors have typically already formed large \HII regions by around $z \sim 10$, together or separately, in many cases trailing by about $\Delta z \sim 1$--$2$ in reaching similar levels of ionization. As discussed earlier, reionization timing is impacted by a variety of factors, which can significantly alter LG ionization histories. For instance, low-mass galaxies near strong ionizing sources can be externally reionized surprisingly early, while isolated massive haloes may reionize later than expected \citep{Dawoodbhoy2018}. As a result, even Milky Way–mass haloes show a wide scatter in reionization redshifts \citep[$z_{\text{reion}} \sim 8$--$15$;][]{Aubert2018}, well before the midpoint of reionization, generally with M31 leading due to its higher mass and more clustered progenitors.

The impact of spatial scale on inferred reionization histories is illustrated in Fig.~\ref{fig:HII_history}, where different line styles correspond to smoothing scales of $\sigma_\text{smooth} = \left\{\text{125\,ckpc, 250\,ckpc, 500\,ckpc, 1\,cMpc}\right\}$. At smaller smoothing scales, the reionization history appears more irregular, reflecting the sensitivity to small-scale structure in the ionizing field and the patchy nature of reionization around individual haloes. Larger smoothing scales yield a more gradual and uniform ionization profile that masks differences between the two haloes.

Fig.~\ref{fig:zreion_hist} shows the distribution of reionization redshifts for haloes in Local Group pairs, separated into the more massive (M31-like) and less massive (MW-like) haloes, evaluated at different smoothing scales. The top panel presents the cumulative distribution functions, while the bottom panel shows the corresponding probability densities. For M31-like haloes, the median $z_\text{reion}$ values with one $\sigma$ uncertainties are $\{13.28^{+1.86}_{-1.85}, 12.99^{+1.38}_{-2.01}, 11.18^{+1.20}_{-1.31}, 9.13^{+1.09}_{-0.83}\}$ for smoothing scales $\sigma_\text{smooth} = \left\{\text{125\,ckpc, 250\,ckpc, 500\,ckpc, 1\,cMpc}\right\}$, respectively. The corresponding values for MW-like haloes are $\{12.71^{+2.04}_{-1.68}, 12.16^{+1.84}_{-1.52}, 10.59^{+1.30}_{-1.17}, 8.88^{+0.66}_{-0.70}\}$. At all smoothing scales, M31-like haloes reionize earlier on average than MW-like haloes. The difference in $z_\text{reion}$ between the two populations is most pronounced at small smoothing scales, where the distributions are broader and more distinct. As the smoothing scale increases, the distributions become later, narrower, and more similar, consistent with reduced sensitivity to local fluctuations in the ionizing radiation field.

Fig.~\ref{fig:dt_histogram} presents the distribution of reionization redshift differences, $\Delta z_\text{reion}$, between the more massive and less massive haloes in each LG pair, evaluated at different smoothing scales. The top panel shows the cumulative distribution function (CDF), while the bottom panel displays the corresponding probability density. The median $\Delta z_\text{reion}$ values with one $\sigma$ uncertainties are $\{0.64^{+2.34}_{-2.79}, 0.38^{+2.61}_{-2.13}, 0.43^{+1.78}_{-1.78}, 0.41^{+1.11}_{-1.07}\}$ for different smoothing scales of $\sigma_\text{smooth} = \left\{\text{125\,ckpc, 250\,ckpc, 500\,ckpc, 1\,cMpc}\right\}$, respectively. These results confirm that, on average, the more massive member of a LG pair reionizes earlier. However, the scatter in $\Delta z_\text{reion}$ highlights the role of local environmental conditions and external ionizing sources in shaping reionization histories. Notably, the scatter increases toward smaller smoothing scales, which are more sensitive to local density fluctuations and source clustering. A detailed analysis of smoothing scale effects is provided in Appendix~\ref{appendix:smoothing}.

The increased width of the $\Delta z_\text{reion}$ distribution at small smoothing scales highlights the significant scatter in reionization timing within LG pairs, reflecting increased sensitivity to local reionization environments. For instance, at $\sigma_\text{smooth} = 125\,\text{ckpc}$, the full range of $\Delta z_\text{reion}$ spans nearly 5 units in redshift. In particular, cases with $\Delta z_\text{reion} < 0$ indicate that the less massive halo reionized earlier than its more massive companion, consistent with the scatter observed throughout this study, especially when affected by early reionization near Virgo-like haloes relative to the larger-scale environment.

At coarse-grained resolution ($\sigma_\text{smooth} = 1\,\text{cMpc}$), the radiation field is averaged over large enough volumes to suppress small-scale fluctuations and reduce the contrast in reionization timing between paired haloes. In contrast, the full simulation resolution captures complex ionization structures such as steep gradients driven by source clustering, butterfly breakout from directional radiative transfer, fluctuations in the ionizing background, and self-shielding effects. The consistently positive median $\Delta z_\text{reion}$ across all smoothing scales, demonstrates the reionization asymmetry between the LG pairs is a robust feature. We interpret this as a systematic preference for earlier reionization correlated with halo mass and overdensity.

\subsection{Correlations between reionization and local properties}
Fig.~\ref{fig:time_sepa_mt_E} presents the dependence of reionization timing in LG pairs on several dynamical properties evaluated at $z = 0$. The top row displays the reionization times, $t_\text{reion}$ of the more and less massive halo in each pair as functions of three parameters: pair separation (left), halo mass ratio (middle), and total energy of the system (right). The bottom row shows the corresponding differences in reionization timing, quantified as $\Delta t_\text{reion} = t_\text{reion}^1 - t_\text{reion}^2$, where superscripts 1 and 2 refer to the more and less massive haloes, respectively.
To illustrate the trends, we divide the sample into five bins for each parameter using the full sample of pairs, with equal numbers of objects. Median values within each bin are shown with error bars in Fig.~\ref{fig:time_sepa_mt_E}. The blue lines represent the best fit to a linear relation, while the shaded bands represent the interquartile range ($25^\text{th}$--$75^\text{th}$ percentile). Individual pairs are plotted as scatter points, with red and green denoting the more and less massive haloes, respectively.

The top-left panel of Fig.~\ref{fig:time_sepa_mt_E} shows no significant correlation between $t_\text{reion}$ and the present-day separation between the two haloes in each LG pair. A constant fit is statistically consistent with the data, indicating that proximity alone does not strongly influence the absolute timing of reionization. However, the median $t_\text{reion}$ values of the more and less massive haloes diverge with increasing separation, suggesting that widely separated pairs tend to reionize at more distinct times. This is confirmed in the bottom-left panel, which shows that pairs with smaller separations tend to exhibit reduced values of $\Delta t_\text{reion}$, often synchronized below 50 Myr. In comparison, more widely separated pairs can have reionization offsets as large as $\sim 150\,\text{Myr}$, with the more massive halo generally reionizing earlier. This is consistent with the conclusion of \citet{Sorce2022}, who found that being in a pair does not systematically lead to earlier reionization.

We define the mass ratio in each LG pair as $M_\text{halo}^2/M_\text{halo}^1$, where $M_\text{halo}^1$ and $M_\text{halo}^2$ denote the mass of the more and less massive haloes, respectively. The dependence of $t_\text{reion}$ and $\Delta t_\text{reion}$ on the mass ratio exhibits a non-monotonic behaviour.
Although the scatter is quite large, there is a small trend for nearly symmetric systems to have reionized earlier and have larger timing offsets. However, the overall variation across bins remains modest, and the scatter is substantial. This suggests that, while a weak correlation may be present, mass ratio is not the primary factor governing reionization time offsets, at least within the mass range probed by our sample of LG pairs.

We then define the total energy of each LG pair as
\begin{equation}
  E = KE + PE = \frac{1}{2} \frac{(v_1 - v_2)^2}{1/M_\text{halo}^1 + 1/M_\text{halo}^2} - \frac{G M_\text{halo}^1 M_\text{halo}^2}{r} \, ,
\end{equation}
where $KE$ and $PE$ denote the kinetic and potential energies of the system, respectively. The kinetic energy depends on both the halo masses and their relative velocity, while the potential energy is determined by the halo masses and their separation. Although total energy is not strictly conserved, it remains a useful proxy for the present-day dynamical state of the pair under the assumption of isolation.
In the top-right panel of Fig.~\ref{fig:time_sepa_mt_E}, we observe a weak negative trend between total energy and $t_\text{reion}$, with more tightly bound systems (i.e., lower total energy) tending to reionize earlier and have more synchronized reionization histories. While the trend is modest and subject to scatter, it is consistent with the expectation that earlier-forming, gravitationally bound systems reionize earlier \citep{Sorce2022}. Altogether these trends are likely driven by whether pairs shared similar assembly environments, which could be teased out with the statistical power of larger volume simulations.

\section{Summary and Conclusions}
\label{sec:conclusions}

In this work, we investigated the reionization histories of Local Group (LG) analogues using the radiation-hydrodynamic simulation \thesanone and its dark matter-only counterpart \thesandarkone. We analyzed how reionization timing across LG analogues depends on halo mass, large-scale environment, and spatial proximity to Virgo-like haloes. By constructing statistically controlled samples of LG pairs and Virgo-like haloes, we explored spatial variations in $z_\text{reion}$, its correlation with environment, and the reionization time offsets between paired haloes across cosmic time. We summarize our main results in the following points:
\begin{enumerate}
  \item In standard reionization models, overdense regions reionize earlier due to enhanced photon production from clustered sources. We confirm this trend by showing that, even at fixed halo mass, haloes in higher local overdensity reionize earlier than those in underdense environments. This correlation is strongest at $z = 5.5$ and weakens at lower redshift, reflecting the nonlinear evolution of large-scale structure. We also find that increasing the smoothing scale of the overdensity field reduces the gradient in $z_\text{reion}$, as environmental differences are averaged out.

  \item We investigate how the choice of smoothing scale in defining overdensity affects the inferred correlation with reionization timing. Using smoothing scales $\sigma_\text{smoothing}$ from 125 ckpc to 1 cMpc, we find that larger smoothing suppresses local fluctuations and reduces the gradient in $z_\text{reion}$, while smaller smoothing scales retain patchy structures and lead to greater scatter.

  \item Massive clusters such as Virgo are expected to host early ionizing sources and influence reionization beyond their virial radii. We identify 20 Virgo-like haloes with $M_\text{halo} (z = 0) > 10^{14}\,\Msun$ in the \thesandarkone simulation, and find that they drive earlier reionization out to $\sim$5--10 cMpc. Outside this range, $z_\text{reion}$ is primarily governed by local sources and overdensities, consistent with a transition from inside-out to external reionization.

  \item We identify 224 LG pairs located within 20\,cMpc of Virgo-like haloes in the \thesandarkone simulation. We find that on average the more massive member reionizes earlier, albeit with large scatter. The difference in reionization timing correlates with pair separation: pairs with small separations ($\lesssim$1 cMpc) tend to reionize nearly simultaneously, while more widely separated systems can exhibit timing offsets up to $\sim$150\,Myr. However, we find no evidence that being in a pair systematically leads to earlier reionization; the timing and offsets are instead primarily governed by local environment and halo properties.
\end{enumerate}

In conclusion, this work establishes a physically consistent framework for understanding the spatial and temporal variations in reionization timing with a particular focus on the Local Group. By combining fully coupled radiation-hydrodynamic and dark matter-only simulations, we identify how halo mass, environmental density, and proximity to massive structures jointly shape reionization histories in both isolated and clustered regions. Building on this framework, future work can investigate how variations in reionization timing affect the subsequent evolution of low-mass galaxies, particularly satellites within LG analogues.

\section*{Acknowledgements}
%
% We thank the referee for a constructive review that greatly improved this work.
RK acknowledges support of the Natural Sciences and Engineering Research Council of Canada (NSERC) through a Discovery Grant and a Discovery Launch Supplement (funding reference numbers RGPIN-2024-06222 and DGECR-2024-00144) and the support of the York University Global Research Excellence Initiative. HL is supported by the National Key R\&D Program of China No. 2023YFB3002502, the National Natural Science Foundation of China under No. 12373006, and the China Manned Space Program with grant No. CMS-CSST-2025-A10. MV acknowledges support through National Science Foundation (NSF) grants AST-2007355 and AST-2107724.

% The Acknowledgements section is not numbered. Here you can thank helpful colleagues, acknowledge funding agencies, telescopes and facilities used etc.

%%%%%%%%%%%%%%%%%%%%%%%%%%%%%%%%%%%%%%%%%%%%%%%%%%
\section*{Data Availability}

All data produced within the \thesan project are fully and openly available at \url{https://thesan-project.com}, including extensive documentation and usage examples \citep{Garaldi2024}. We invite inquiries and collaboration requests from the community.
 
% The inclusion of a Data Availability Statement is a requirement for articles published in MNRAS. Data Availability Statements provide a standardised format for readers to understand the availability of data underlying the research results described in the article. The statement may refer to original data generated in the course of the study or to third-party data analysed in the article. The statement should describe and provide means of access, where possible, by linking to the data or providing the required accession numbers for the relevant databases or DOIs.



\bibliographystyle{mnras}
\bibliography{biblio}

%%%%%%%%%%%%%%%%%%%%%%%%%%%%%%%%%%%%%%%%%%%%%%%%%%

%%%%%%%%%%%%%%%%% APPENDICES %%%%%%%%%%%%%%%%%%%%%

\appendix

\section{Bijection Fraction}
\label{appendix:bijection}
\begin{figure}
\includegraphics[width=\columnwidth]{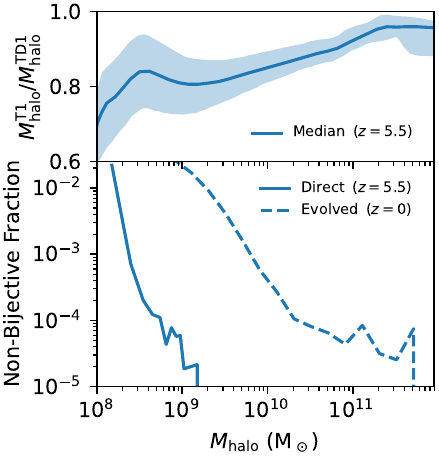}
  \caption{Top: Median ratio of subhalo masses in \thesanone and \thesandarkone at $z = 5.5$, showing $M_\mathrm{halo}^\mathrm{T1}/M_\mathrm{halo}^\mathrm{TD1}$ as a function of $M_\mathrm{halo}$ in \thesanone. Bottom: Non-bijective fraction as a function of halo mass at $z = 5.5$, defined as the number of haloes in \thesanone that do not have a bijective match to \thesandarkone, divided by the total number of subhaloes in the same mass bin. The solid curve shows the direct mismatch at $z = 5.5$, while the dashed curve shows the evolved masses of these object in \thesandarkone at $z = 0$. The fraction decreases to zero above $\approx 2 \times 10^9\,\Msun$ in \thesanone.}
  \label{fig:bijection}
\end{figure}

We briefly demonstrate the effectiveness of the bijection fraction for dark-matter-only to hydrodynamical simulated haloes, defined as the number of \thesanone subhaloes that have a one-to-one match in \thesandarkone, divided by the total number of subhaloes in each mass bin. In the top panel of Fig.~\ref{fig:bijection}, we show the median ratio between the halo masses in \thesanone and their bijective matches in \thesandarkone, with shaded regions indicating the $16^\mathrm{th}$ to $84^\mathrm{th}$ percentile ranges. In the bottom panel, we plot the non-bijective fraction, which drops to zero by a halo mass of $\approx 2 \times 10^9\,\Msun$ at $z=5.5$. The solid line shows the result for \thesanone haloes at $z = 5.5$, while the dashed line shows the corresponding result for \thesandarkone haloes at $z = 0$. Excluding low-mass unmatched haloes has a minimal impact on our analysis, as the hierarchical structure is well-matched but detailed properties might not be.

\section{IMPACT OF GRID SMOOTHING SCALE}
\label{appendix:smoothing}

\begin{figure*}
\includegraphics[width=\textwidth]{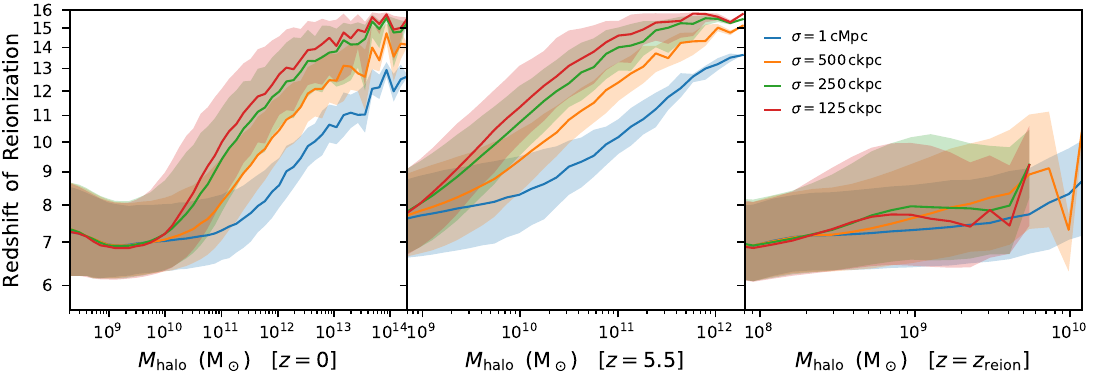}
 \caption{Correlation between reionization redshift $z_\text{reion}$ and halo mass $M_\text{halo}$ at $z = 0$ (left panel), $z = 5.5$ (middle panel), and at the individual reionization redshift of each halo $z = z_\text{reion}$ (right panel), comparing Gaussian smoothing scales $\sigma_\text{smooth} = \left\{\text{125\,ckpc, 250\,ckpc, 500\,ckpc, 1\,cMpc}\right\}$. The solid lines show the median values, and the shaded regions indicate the $16^\text{th}$ to $84^\text{th}$ percentiles. Smaller smoothing scales yield earlier $z_\text{reion}$ at fixed halo mass, with the largest differences seen at $z = 5.5$. By $z = 0$, the curves converge across smoothing scales. At $z = z_\text{reion}$, both mass dependence and smoothing effects are weak.}
\label{fig:mzreionsmoothing}
\end{figure*}

\begin{figure*}
\includegraphics[width=\textwidth]{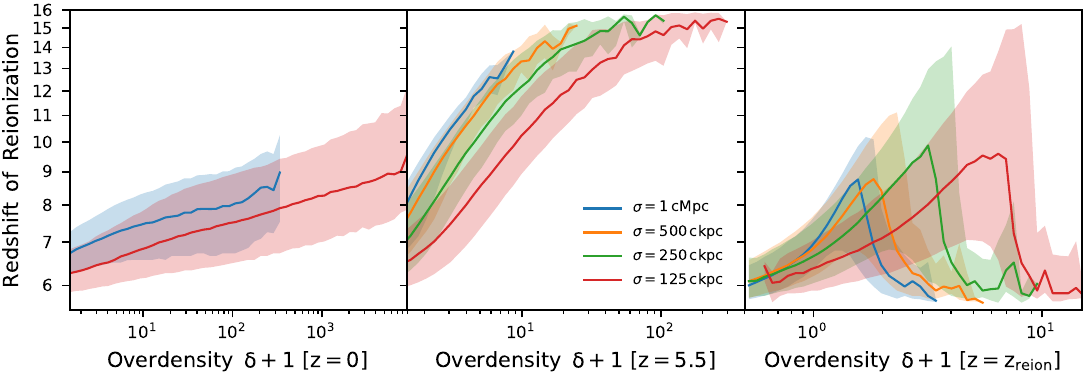}
 \caption{Similar to Fig.~\ref{fig:mzreionsmoothing} but for local overdensity. Differences across smoothing scales are most visible at $z = 5.5$ at low to intermediate overdensity.}
\label{fig:overdzreionsmoothing}
\end{figure*}

We investigate how the choice of Gaussian smoothing scale $\sigma_\text{smooth}$ affects the correlation between reionization redshift and halo properties. This procedure suppresses small-scale variations while preserving large-scale environmental structure relevant for reionization. Fig.~\ref{fig:mzreionsmoothing} shows the dependence of the $z_\text{reion}$--$M_\text{halo}$ relation on the choice of Gaussian smoothing scale, evaluated at $z = 0$ (left), $z = 5.5$ (middle), and at the reionization redshift of each halo $z = z_\text{reion}$ (right). The solid lines indicate the median $z_\text{reion}$ as a function of halo mass, and the shaded regions show the $16^\text{th}$ to $84^\text{th}$ percentiles. Curves are shown for smoothing scales of $\sigma_\text{smooth} = \left\{\text{125\,ckpc, 250\,ckpc, 500\,ckpc, 1\,cMpc}\right\}$.

Across all smoothing scales, there is a strong positive correlation between $z_\text{reion}$ and $M_\text{halo}$ at $z = 5.5$ and $z = 0$, especially for haloes with $M_\text{halo} \gtrsim 10^{10}\,\Msun$. This trend is consistent with earlier reionization in overdense regions. At $z = z_\text{reion}$, the correlation is weaker due to the narrower mass range and the presence of low-mass haloes reionized by external sources. Smaller smoothing scales yield earlier median $z_\text{reion}$ at fixed halo mass. This difference arises from the enhanced sensitivity to local overdensities when using finer smoothing. The offset is largest at $z = 5.5$, where the median values differ significantly for the overly-aggressive 1\,cMpc smoothing. By $z = 0$, the impact of smoothing is reduced. Median trends start to converge across scales, indicating that late-time mass growth diminishes the imprint of early environmental differences. At $z = z_\text{reion}$, the results are similar across smoothing scales, stacking similar halo masses over a wide redshift range, biased in number by the late-reionized lower-mass haloes.

Fig.~\ref{fig:overdzreionsmoothing} shows the dependence of the $z_\text{reion}$--($\delta + 1$) relation on Gaussian smoothing scale, evaluated at $z = 0$ (left), $z = 5.5$ (middle), and $z = z_\text{reion}$ (right). At $z = 5.5$, the curves display the largest separation, with smaller smoothing scales yielding earlier reionization at fixed $\delta$. The offset is most prominent at low to intermediate overdensities, and the ordering of curves is monotonic across all scales. At $z = 0$, the trend persists but the separation narrows, especially at high $\delta$, indicating reduced sensitivity to the choice of $\sigma_\text{smooth}$. At $z = z_\text{reion}$, the median trends remain ordered by smoothing scale, but the curves diverge in shape and show increased scatter. The peak locations shift with $\sigma_\text{smooth}$, and low-overdensity haloes span a wide range in reionization redshifts. The persistence of scale-dependent differences across all panels highlights the impact of early-time density fluctuations and their sensitivity to the smoothing prescription.

\section{IMPACT OF IONIZED-FRACTION THRESHOLD}
\label{appendix:HII-threshold}

\begin{figure*}	\includegraphics[width=\textwidth]{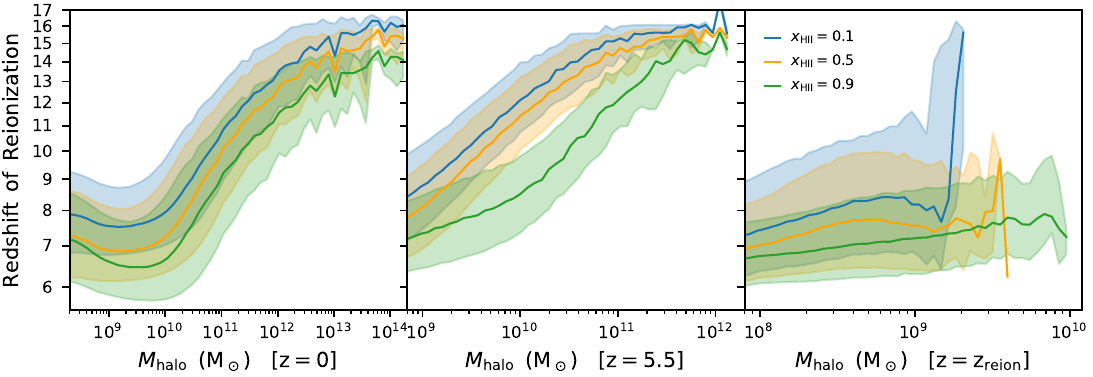}
 \caption{Relation between reionization redshift and subhalo mass at $z = \left\{0, 5.5, z_\text{reion}\right\}$, evaluated using three ionized-fraction thresholds $x_{\HII} = \left\{0.1, 0.5, 0.9\right\}$. Solid lines show the median $z_\text{reion}$ values in each mass bin, and shaded regions indicate the $16^\text{th}$--$84^\text{th}$ percentile range. Different colors correspond to different threshold choices, illustrating the sensitivity of inferred reionization redshifts to the adopted ionization criterion. At all redshifts, increasing the ionization threshold leads to later inferred reionization redshifts. The positive correlation with halo mass is most pronounced at $z = 5.5$.}
 \label{fig:mzreionxHI}
\end{figure*}

\begin{figure*}	\includegraphics[width=\textwidth]{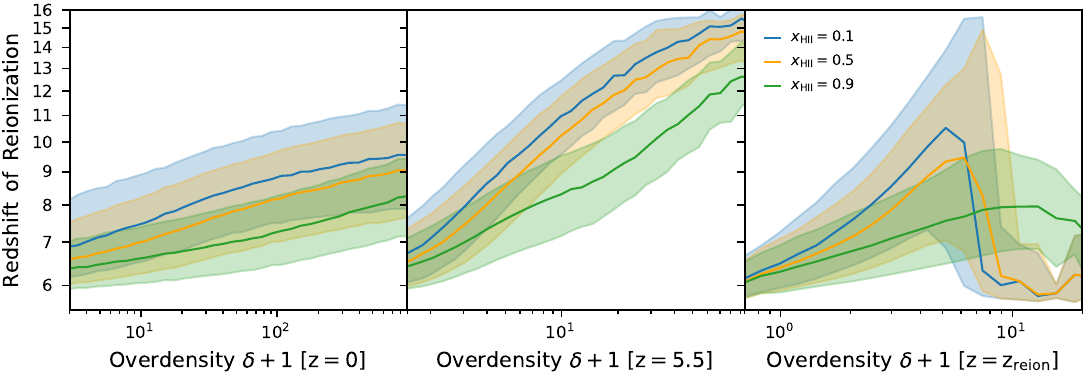}
 \caption{Similar to Fig.~\ref{fig:mzreionxHI} but for local overdensity. The choice of threshold leads to systematic biases but no qualitative changes.}
 \label{fig:overdzreionxHI}
\end{figure*}

Throughout this work, we have defined $z_\text{reion}$ as the redshift corresponding to the earliest epoch at which the volume-averaged ionized hydrogen fraction $x_\text{HII}$ is persistently above 0.5. Here, we explore how the inferred reionization history depends on the choice of threshold by repeating the analysis with alternative values of $x_{\HII} = \{0.1, 0.5, 0.9\}$. These thresholds correspond to different stages of the reionization process: $x_{\HII} = 0.1$ indicates the early onset of ionization, $x_{\HII} = 0.5$ marks the midpoint, and $x_{\HII} = 0.9$ reflects the near-completion stage, when most of the gas is ionized.

Fig.~\ref{fig:mzreionxHI} explores how the choice of ionization threshold affects the inferred relation between halo mass and reionization redshift at three representative redshifts, $z = \{0, 5.5, z_\text{reion}\}$, and ionization thresholds $x_{\HII} = 0.1$, $0.5$. Across all redshifts and thresholds, more massive haloes tend to reionize earlier, but the absolute values of $z_{\rm reion}$ shift systematically with threshold choice. These offsets are most pronounced at $z = 0$ and $z = 5.5$, where the time interval between initial and final ionization is relatively long. In contrast, at $z = z_\text{reion}$, the mass dependence is weaker and the three curves are more closely aligned, reflecting the more synchronized reionization of low-mass haloes at later epochs.

Fig.~\ref{fig:overdzreionxHI} examines the sensitivity to the choice of ionization threshold but for the $z_\text{reion}$--$(\delta + 1)$ correlation, comparing the fiducial value of $x_{\HII}=0.5$ to earlier ($x_{\HII}=0.1$) and later ($x_{\HII}=0.9$) stages of the reionization process. Across all redshifts, the qualitative trend remains: haloes residing in more overdense regions tend to reionize earlier, as expected by source clustering and accelerated local ionization. Quantitative differences arise, however, depending on the threshold adopted. At fixed overdensity, increasing the ionization threshold leads to later $z_\text{reion}$. The largest differences are observed in moderately overdense environments, where the spread in reionization timing is greatest. By contrast, in low-density regions, the three curves converge, indicating a rapid and late synchronized reionization. Although the rank ordering of haloes by reionization timing is largely preserved, the inferred values of $z_{\rm reion}$ depend on the ionization criterion.

%%%%%%%%%%%%%%%%%%%%%%%%%%%%%%%%%%%%%%%%%%%%%%%%%%

% Don't change these lines
% \bsp	% typesetting comment
\label{lastpage}
\end{document}